# Cross-scale Modeling of Polymer Topology Impact on Extrudability through Molecular Dynamics and Computational Fluid Dynamics

*Yawei Gao[a], Jan Michael Carrillo[b], Logan T. Kearney[a], Polyxeni P. Angelopoulou[a], Nihal Kanbargi[a], Arit Das[a], Michael Toomey[a], Bobby G. Sumpter[b], Joshua T. Damron[a*], and Amit K. Naskar[a*]*

[a]Chemical Sciences Division, Oak Ridge National Laboratory, 1 Bethel Valley Rd, Oak Ridge, TN 37830

[b]Center for Nanophase Materials Sciences, Oak Ridge National Laboratory, 1 Bethel Valley Rd, Oak Ridge, TN 37830

**Abstract**

Understanding how polymer topology influences melt extrudability is critical for advancing material design in extrusion-based additive manufacturing. In this work, we develop a bottom-up, cross-scale modeling framework that integrates coarse-grained molecular dynamics (CGMD) and continuum-scale computational fluid dynamics (CFD) to quantitatively assess the effects of polymer architecture on extrudability A range of branched polydimethylsiloxane (PDMS) polymers are systematically designed by varying backbone length, sidechain length, grafting density, grafted block ratio, and periodicity of grafted-ungrafted segments. CGMD simulations are used to compute zero-shear viscosity and relaxation times, which are then incorporated into the Phan-Thien-Tanner (PTT) model within a computational fluid dynamics (CFD) model to predict pressure drop of PDMS during extrusion through printer nozzle. Qualitative analysis reveals that  polymers with concentrated grafted blocks exhibit significantly higher zero-shear viscosity than stochastically branched analogs, while sidechain inertia drives longer relaxation time. However, for untangled and weakly entangled PDMS, relaxation time remains in the nanosecond range, making shear-thinning and elastic effects negligible. Consequently, zero-shear viscosity emerges as the primary determinant of extrudability. This cross-scale modeling strategy provides a predictive framework for guiding the rational design of extrudable polymer materials with tailored topologies.

[*] Corresponding author.

*Email addresses*: damronjt@ornl.gov (Joshua T. Damron),  naskarak@ornl.gov (Amit. K. Naskar)

*May 1, 2025*

## 1. Introduction

Recent advancements in polymerization techniques have enabled the synthesis of polymers with diverse topological attributes, including variations in graft density, side-chain length, graft heterogeneity, and backbone length.[1] In our recent study,[2] we demonstrated that tuning polymer topology can significantly improve interfacial welding quality in fused filament fabrication(FFF), highlighting topological design as a promising lever for optimizing additive manufacturing performance.[2] However, broader implementation of topologically complex polymers in FFF remains limited by challenges in melt printability.[3,4] Previous studies have explored various rheological models to better understand and predict melt printability. For example, Ngoc et al. used a power-law-based model to investigate the printability of sustainable composites.[5] Although their model was effective at capturing shear-thinning behavior, it overlooks elastic effects, resulting in inaccuracies under high shear-rate flows, often encountered by the polymer melt at the nozzle during FFF. In contrast, the Phan-Thien-Tanner (PTT) model provides a more comprehensive viscoelastic description by incorporating both shear-thinning and elasticity effects via the pressure tensor.[6] Still, translating these macroscopic rheological parameters - such as zero-shear viscosity and relaxation time - into actionable guidance for polymer synthesis is nontrivial, as these properties are often interdependent. For instance, Lohse et al. studied the rheological behavior of polyethylene (PE) melts with various topologies, including linear, H-shaped, pom-pom, and comb-like architectures. They found that star and comb polymers with 30 branches exhibited higher zero-shear viscosity than linear chains of greater molecular weight, and that longer sidechains amplified shear-thinning behavior.[7]

Coarse-grained molecular dynamics (CGMD) simulations offer a more direct approach to engineering polymer topology for targeted rheological properties. Several simulation studies have explored the interplay between topology and rheology. Mukkamala et al. employed dissipative particle dynamics (DPD) based CGMD in combination with proper orthogonal decomposition (POD) to investigate the relaxation dynamics of bottlebrush polymers in dilute polymer solution. They found that while the backbones behave like linear chains, side chains heavily influenced monomer motion.[8] Similarly, Wijesinghe et al. used CGMD to examine the impact of topology on chain mobility and viscosity in comb-like PE systems, concluding that side-chain length, rather than number, played a dominant role in determining rheological behavior.[9] Additionally, CGMD based predictions provide valuable insights into the structural and mechanical properties by offering molecular-level design rules to enable 3D printing of soft matter.[10] Previous work using MD simulations has also indicated that by manipulating the molecular architecture specifically the grafting density of bottlebrush polymers - the shear thinning behavior can be tuned to optimize flow during printing.[11] Specifically, for the case of FFF, "printability" refers to a polymer's ability to be consistently extruded and deposited into accurate, defect-free geometries. It is quantitatively assessed using metrics such as geometric fidelity, flow consistency, and shape retention.[3, 4] For our current work, we primarily focus on the first requirement of printability, namely, extrudability of a viscoelastic polymer melt through the printer nozzle.

The primary objective of the current manuscript is to leverage insights from MD simulations to advance the design of extrudable viscoelastic polymers for FFF? To bridge molecular-scale topology with macroscopic extrusion behavior, we propose a bottom-up cross-scale simulation strategy that combines CGMD with PTT-based CFD simulations (**Figure 1**). In this approach, CGMD simulations are used to determine the zero-shear viscosity and relaxation time of polymers with different topologies. These rheological properties are then incorporated into PTT-based CFD simulations to evaluate extrudability, quantified by the pressure drop during flow. A higher pressure drop indicates the need for greater inlet pressure to maintain a fixed printing rate, suggesting reduced extrudability of the polymer melt.

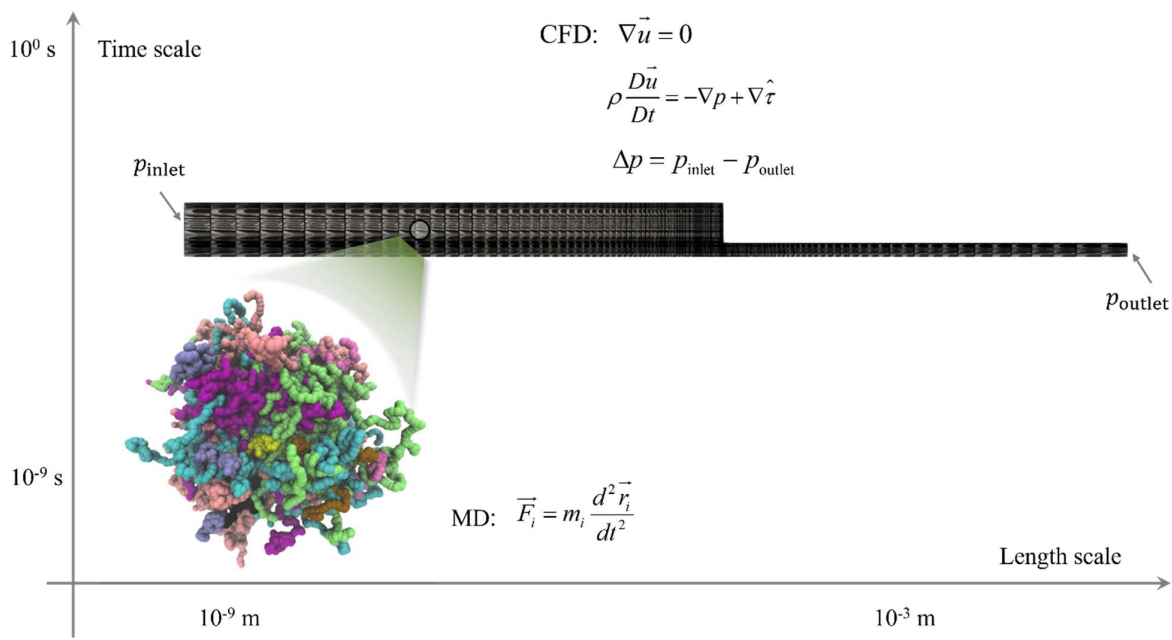

Figure 1: Cross-scale modeling strategy linking polymer topology to viscoelastic material printability. We use CFD to investigate the pressure drop of molten polymer flowing through a nozzle ($\Delta p = p_{inlet} - p_{outlet}$). The viscoelastic rheology for CFD simulations is derived from CG MD simulations.

We apply this approach to a library of hierarchically designed polymer melts, varying in both conventional topological parameters - such as backbone length, graft density, and side-chain length - and two novel architectural parameters aimed at capturing heterogeneity:

1. **Graft block length ratio**: A 50% ratio implies equal lengths of grafted and ungrafted segments.

2. **Number of grafted-ungrafted periods**: For example, a polymer with four grafted (G) and four ungrafted (U) segments can be arranged as one period (GGGGUUUU) or four alternating periods (GUGUGUGU).

At the molecular level, we compute zero-shear viscosity using both the Green-Kubo and Einstein relations and analyze relaxation dynamics via Rouse mode decomposition. These parameters are then fed into the CFD model to evaluate printability in terms of pressure drop across the nozzle. We believe this cross-scale strategy offers a robust framework for guiding the rational design and synthesis of high-performance, extrudable viscoelastic materials for FFF applications.

## 2. Simulation Details
2.1. Polymer Topology Designs

While the conformational freedom arising from both backbone and sidechain structures can be broadly tuned, key questions remain regarding how polymer architecture influences relaxation dynamics. To address this, we propose a hierarchical strategy for systematically controlling the topology of blocky and periodically grafted polymers. As illustrated in **Figure 2**, in addition to manipulating backbone length, sidechain length, and grafting density, we also introduce two topological descriptors to capture architectural heterogeneity: the parameter $n_p$, which governs the spatial distribution of branches, and the composition factor $\varphi = n_{graft}/N_{bb} = n_{graft}/(n_{graft} + n_{ungraft})$, which defines the relative length of grafted versus ungrafted blocks along the backbone. In this context, $n_{graft}$ and $n_{ungraft}$ represent the lengths of the grafted and ungrafted blocks on the backbone, respectively. This design enables the construction of branched polymers with linearly connected backbones and topologically heterogeneous architectures.

We selected backbone lengths of $N_{bb} = 80$ and 240, corresponding to approximately one and three entanglement lengths of flexible backbone, respectively, to represent unentangled and weakly entangled polymers ($N_e \approx 85 - 90$ for fully flexible chain polymers in Ref.[12]). We set the grafted-ungrafted block periods to 1, 2, and 4, to investigate whether a concentrated or a dispersed grafting pattern is more favorable for optimizing polymer printability. In addition to the case of $\varphi = 0.5$, which exhibits strong architectural heterogeneity (equal grafted and ungrafted block lengths), we also included a mild heterogeneous configuration with $\varphi = 0.1$, which indicates that 90% of the backbone consists of ungrafted blocks. The minimum graft density $\sigma_{sc}$ was set to 25% to ensure that the all grafted segments fell beyond the loosely grafted comb regime.[13] Finally, sidechain lengths $N_{sc} = 4$ and 12 were chosen to remain significantly shorter than the backbone, preserving the branched-linear polymer classification with a clearly distinguishable backbone.

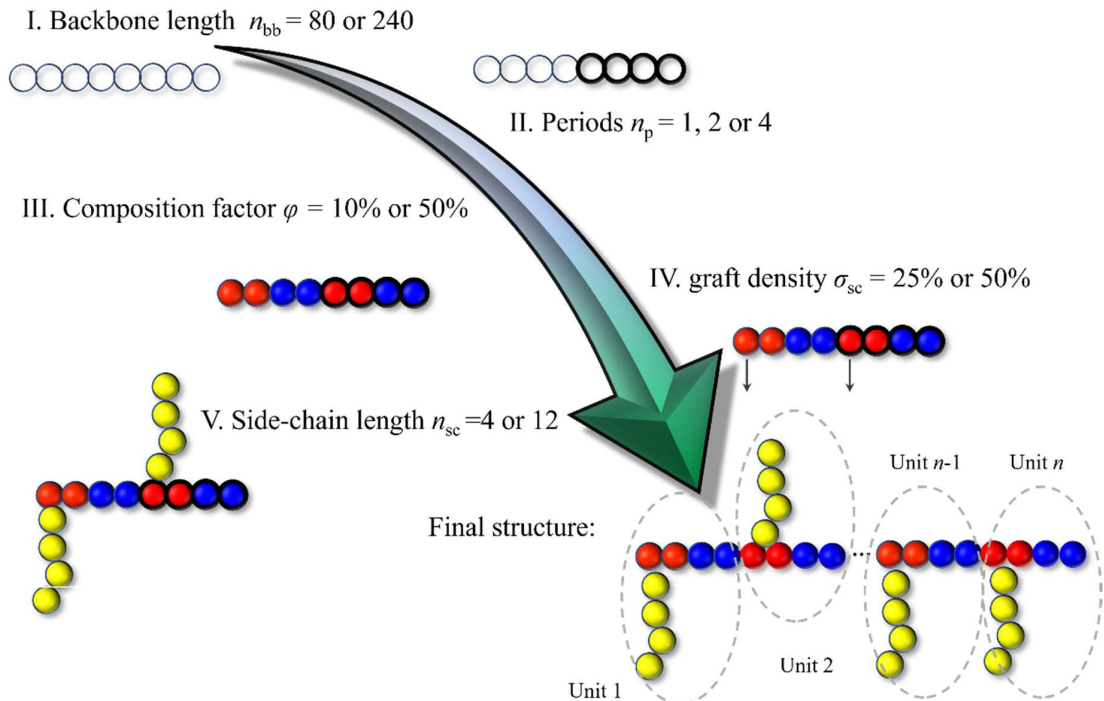

Figure 2: A hierarchical strategy to build a blocky and periodically grafted polymer from the bottomup. In the schematic, red and blue beads in (III, IV and V) refer to the grafted and ungrafted blocks in the backbone, respectively, while the yellow beads denote the sidechains. A grafted-ungrafted unit (in II, III, IV and V) is outlined in black. This example illustrates the modulation of the backbone length $N_{bb}$, the number of periods $n_p$, composition factor $\varphi = n_{graft}/(n_{graft} + n_{ungraft})$, graft density, and sidechain length.

However, performing a full factorial design of simulations to explore all five factors in **Figure 2** is computationally prohibitive, requiring 48 (2 × 2 × 2 × 2 × 3) combinations. To address this issue, we adopt a Taguchi method based fractional factorial design, which offers a more practical and cost-effective approach. By employing orthogonal pairing, this method strategically omits certain combinations while preserving statistical balance across factors. As noted in Ref.[14], orthogonal pairing can reduce the required number of simulation cases from 48 to just 12 for one 3-level factor and four 2-level factors, as outlined in **Table 1**.

Table 1: Taguchi design for CGMD simulations. The topological structures of each polymer can be visualized in **Figure S1**.

| | label | backbone length $N_{bb}$ | periods $n_p$ | Composition factor $\varphi$ | graft density $\sigma_{sc}$ | Sidechain length $N_{sc}$ |
|---|---|---|---|---|---|---|
| Case 01 | 80p4φ50σ50s4 | 80 | 4 | 50% | 50% | 4 |
| Case 02 | 80p2φ10σ50s4 | 80 | 2 | 10% | 50% | 4 |
| Case 03 | 80p2φ50σ50s12 | 80 | 2 | 50% | 50% | 12 |
| Case 04 | 240p2φ50σ25s12 | 240 | 2 | 50% | 25% | 12 |
| Case 05 | 240p2φ10σ50s4 | 240 | 2 | 10% | 50% | 4 |
| Case 06 | 240p4φ10σ25s4 | 240 | 4 | 10% | 25% | 4 |
| Case 07 | 240p4φ50σ25s12 | 240 | 4 | 50% | 25% | 12 |
| Case 08 | 240p1φ50σ50s4 | 240 | 1 | 50% | 50% | 4 |
| Case 09 | 80p1φ50σ25s4 | 80 | 1 | 50% | 25% | 4 |
| Case 10 | 80p1φ10σ25s12 | 80 | 1 | 10% | 25% | 12 |
| Case 11 | 80p4φ10σ50s12 | 80 | 4 | 10% | 50% | 12 |
| Case 12 | 240p1φ10σ50s12 | 240 | 1 | 10% | 50% | 12 |

## 2.2. Simulation Methods

### 2.2.1. CGMD Formalism

All polymers are modeled with the generic Kremer-Grest (KG) spring-bead model. The CGMD simulations are carried out on the SUMMIT and FRONTIER supercomputers,[15] employing GPU acceleration implemented in Large-scale Atomic/Molecular Massively Parallel Simulation (LAMMPS).[16] To maintain generality in our CGMD simulations, all quantities are in Lennard-Jones (LJ) dimensionless units, scaled by fundamental properties: mass, distance ($\sigma$), time ($\tau$), energy ($\epsilon$), and the Boltzmann constant $k_B = 1$. Moreover, to decouple polymer topology from chemical specifics, the coarse-grained (CG) beads representing the grafted blocks, ungrafted blocks, and sidechains are assigned identical interaction parameters. Within the KG model scheme, the bonded interaction is modeled with the finite-extensible-nonlinear (FENE) potential plus the 12-6 Lennard-Jones (LJ) potential with $1.0\sigma$ and $1.0\epsilon$, as given in **Equation 1**:

$$U_{\text{FENE}}(r) = -0.5 K_r R_0^2 \ln\left[1 - \left(\frac{r}{R_0}\right)^2\right] + 4\epsilon\left[\left(\frac{\sigma}{r}\right)^{12} - \left(\frac{\sigma}{r}\right)^6\right] + \epsilon, \quad (1)$$

where $r$ is the distance between bonded CG beads. $K_r$ and $R_0$ are FENE potential parameters that are specified to $30.0\epsilon$ and $1.5\sigma$, respectively. Furthermore, unless otherwise noted, a flexible backbone and sidechains are employed without any constraint on the bending rigidity.

Non-bonded interactions between CG beads are described by a shifted and truncated LJ potential with a cutoff distance of $2.5\sigma$. Each polymer architecture detailed in **Table 1** is used to construct a neat melt, with systems comprising approximately 10,000 CG beads. Initially, polymer chains were randomly dispersed within a cubic simulation box with periodic boundary conditions at a density of $0.85/\sigma^3$. To relax initial overlaps, bead displacements were restricted to a maximum of $0.01\sigma$ per timestep. This was followed by equilibration under an isothermal-isobaric (NPT) ensemble at zero pressure. After NPT-MD equilibration, simulations proceed in canonical (NVT) ensemble, using a Langevin thermostat with a damping constant $\Gamma = 1.0\tau^{-1}$ at temperature $T^* = 1.0$ with a timestep of $0.01\tau$.

*2.2.2. Green-Kubo and Einstein Methods*

One of the crucial rheological parameters of the collective dynamics of polymer melts is the zero-shear viscosity $\eta_0$, which can be determined from several approaches.[17-24] To enhance the accuracy of viscosity calculation, we utilized both the Green-Kubo and the Einstein relationships.

In the Green-Kubo equation,[25]

$$\eta_0 = \lim_{t \to \infty} \frac{V}{k_B T} \int_0^t \langle \sigma_{ij}(t)\sigma_{ij}(0)\rangle dt \tag{2}$$

where $V$, $k_B$, and $T$ are simulation box volume, Boltzmann constant, and temperature, respectively. $\langle \tau_{ij}(t)\tau_{ij}(0)\rangle$ refers to the stress autocorrelation function $G(t)$ averaged over all possible pairs, as shown in **Equation 3**.[26]

$$\begin{aligned}G(t) = &\frac{1}{5}\left[\langle\sigma_{xy}(t)\sigma_{xy}(0)\rangle + \langle\sigma_{xz}(t)\sigma_{xz}(0)\rangle + \langle\sigma_{yz}(t)\sigma_{yz}(0)\rangle\right] \\ &+ \frac{1}{30}\left[\langle N_{xy}(t)N_{xy}(0)\rangle + \langle N_{xz}(t)N_{xz}(0)\rangle + \langle N_{yz}(t)N_{yz}(0)\rangle\right]\end{aligned} \tag{3}$$

with $N_{ij} = \sigma_{ii} - \sigma_{jj}$. Here, $\sigma_{ii}$ and $\sigma_{jj}$ denote $\sigma_{xx}$, $\sigma_{yy}$, and $\sigma_{zz}$.

An alternative method for viscosity calculation is based on the Einstein relation, in which $\eta$ is stated in Eq. 4[27]

$$\eta_0 = \lim_{t \to \infty} \frac{V}{2tk_B T}\langle\left(\int_0^t \sigma_{ij}(t')dt'\right)^2\rangle \tag{4}$$

It is well-known that calculating viscosity for polymer melts can be challenging due to persistent noise in pressure tensors. To address this challenge, the integral was averaged over at least 80 MD trajectories for each polymer melt configuration to reduce noise. Subsequently, the zero-shear viscosity $\eta_0$ was modeled with a double-exponential function as **Equation 5**.[24]

$$\eta(t) = \eta_0 \alpha \left(1 - e^{-\frac{t}{\lambda_1}}\right) + \eta_0 (1 - \alpha) \left(1 - e^{-\frac{t}{\lambda_2}}\right), \tag{5}$$

where $\eta_0$, $\alpha$, $\lambda_1$, and $\lambda_2$ are parameters to fit. As $t \to \infty$, $\eta(t) = \eta_0$.

### 2.2.3. Rouse Mode Analysis

Though the Rouse model for polymer melts is originally designed for unentangled linear polymer relaxation analysis,[28] it has already been proven applicable for long polymers.[12] Besides, Ref.[8] pointed out that the relaxation of the entire branched polymer for $N_{bb} > 50$ can be approximated with the backbone relaxation. Therefore, we performed Rouse mode analysis (RMA) to quantify the influence of topology on relaxation times. The RMA is based on the Rouse mode coordinates $X_p$ (given in **Equation 6**):[29]

$$\boldsymbol{X}_p = \left(\frac{2}{N}\right)^{1/2} \sum_{i=1}^{N} \boldsymbol{r}_i(t) \cos\left[\frac{p\pi}{N}(i - 1/2)\right] \tag{6}$$

Here, the $p$ modes denote the internal relaxation of $N/p$ segments in a chain. The mode of $p = 0$ corresponds to the center-of-mass motion, while other modes ranging from 1 to $N$-1 describe internal relaxation subsegments proportionate to their length $N/p$. Furthermore, the normalized autocorrelation function of $X_p$ can conform to a stretched exponential Kohlrausch-Williams-Watts (KWW) function,[12, 30-32] i.e.,

$$\frac{\langle \boldsymbol{X}_p(t) \boldsymbol{X}_p(0) \rangle}{\langle \boldsymbol{X}_p(0) \boldsymbol{X}_p(0) \rangle} = e^{-(t/\tau_p)^{\beta_p}} \tag{7}$$

Here, both the stretching exponent $\beta_p$ and the KWW characteristic relaxation time $\tau_p$ reflect the impact of excluded volume interactions and entanglements, which vary with mode $p$ independently. As noted by Ref.[33], the Rouse modes are not the correct normal modes of long polymer melts in the entangled regime, since interchain topological interactions violate the core assumption of interchain decoupling in the Rouse theory. Nonetheless, Rouse mode analysis remains beneficial to interpret the polymer dynamics and informing the design of optimized polymer topology. Previous studies on neat melts have shown that or short, unentangled linear polymers, $\beta_p$ increases from approximately 0.8 to 1.0 as $p$ approaches 1. In contrast, for long, entangled polymers, $\beta_p$ tends to decrease, reaching a minimum near the mode associated with the entanglement length $N_e$.[12, 32, 33] These findings serve as benchmarks for our Rouse mode analysis.

### 2.2.4. Computational fluid dynamics

Given that the nozzle diameter of a typical FFF printer is approximately millimeter scale, modeling the viscoelastic melt flow at this scale exceeds the capabilities of CGMD simulations. Therefore, a continuum-scale CFD approach is performed with an open-source finite-volume method (FVM) solver - *RheoTool*[34-36]. In CFD, mass conservation is ensured by the continuity equation displayed in **Equation 8**:

$$\nabla \cdot \boldsymbol{u} = 0 \tag{8}$$

where $\boldsymbol{u}$ is the velocity vector, and $\nabla \cdot \boldsymbol{u}$ represents the velocity gradient. The momentum conservation is described by the Navier Stokes (NS) equation as shown in **Equation 9**:

$$\rho \frac{D\boldsymbol{u}}{Dt} = -\nabla p + \nabla \cdot \boldsymbol{\sigma} \tag{9}$$

where $\boldsymbol{\sigma}$ refers to viscoelastic stress tensor, and $p$ and $\rho$ denote the pressure and mass density, respectively. The viscoelastic behavior of pressure tensor $\boldsymbol{\sigma}$ is described by a constitutive model based on the PTT model[6]:

$$\exp\left(-\frac{\varepsilon \tau_1}{\eta_0} \operatorname{tr}(\boldsymbol{\sigma})\right) \boldsymbol{\sigma} + \tau_1 \stackrel{\nabla}{\boldsymbol{\sigma}} = 2\eta_0 \boldsymbol{D} \tag{10}$$

The upper-convective time derivative $\stackrel{\nabla}{\boldsymbol{\sigma}}$ is defined in **Equation 11:**

$$\stackrel{\nabla}{\boldsymbol{\sigma}} = \frac{D\boldsymbol{\sigma}}{Dt} - (\nabla \cdot \boldsymbol{u})^{\mathrm{T}} \cdot \boldsymbol{\sigma} + \boldsymbol{\sigma} \cdot (\nabla \cdot \boldsymbol{u}) \tag{11}$$

with $\varepsilon$ being the dimensionless parameters accounting for polymer extensibility, which is set to 0.25 unless otherwise noted.[37-39] The rate-of-deformation tensor $\boldsymbol{D}$ is defined as $\boldsymbol{D} = [\nabla \cdot \boldsymbol{u} + (\nabla \cdot \boldsymbol{u})^{\mathrm{T}}]/2$. The PTT model relies on two key inputs: the zero-shear viscosity $\eta_0$ and polymer relaxation time $\tau_1$, both obtained by the CGMD simulations. **Figure S2** exhibits the two-dimensional flow channel employed in our CFD simulations, designed to capture both shear flow and contraction flow behavior.[34] Boundary conditions are provided in **Table S1**. The extrudability of a polymer melt is quantified by the pressure drop between the inlet and outlet of the nozzle.

### 3. Results and Discussions

*3.1 Zero-shear viscosity*

. To improve the accuracy of zero-shear viscosity in our study, we implemented three measurements. First, we averaged over 80 independent MD trajectories to derive $\eta(t)$, exceeding the typical requirement of 30 ~ 40 trajectories as suggested in Ref.[24] Secondly, the resulting $\eta(t)$ data were fitted with a double exponential function (as outlined in **Equation 5**), and $\eta_0$ was extrapolated in the limit as $t \to \infty$. Lastly, we ensured the consistency of zero-shear viscosity derived from the Green-Kubo relation and the Einstein relation before conducting further statistical analyses.

3.1.1 Zero-shear viscosity calculation via Green-Kubo method

As demonstrated in **Equation 3**, the zero-shear viscosity $\eta_0$ calculated using the Green-Kubo integrating the autocorrelation function of the stress tensor, $G(t)$, as it gradually decays to 0.

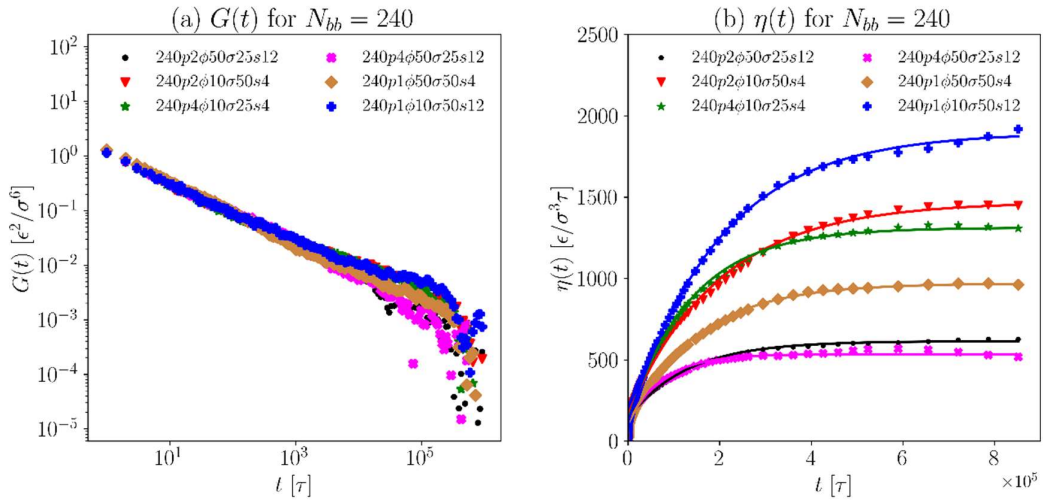

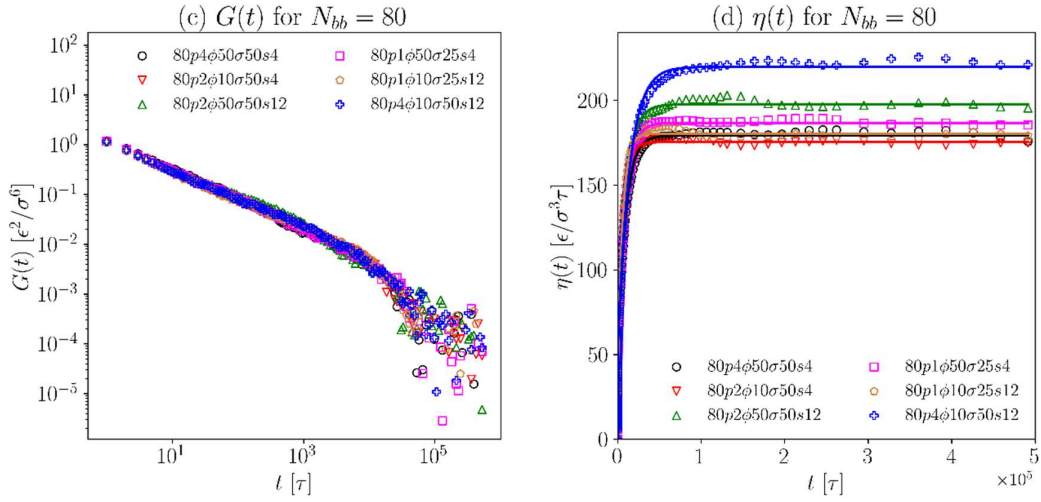

Figure 3: Stress autocorrelation function $G(t)$ (a,c) and zero-shear viscosity calculations $\eta(t)$ (b,d) via the Green-Kubo method in conjunction with the double exponential function (solid lines). The topological details are provided in **Figure 1.**

**Figure 3**(a) and (c) display the stress autocorrelation functions $G(t)$ for all polymer architectures in our CGMD simulations. Based on the decay behavior, integration time of $5 \times 10^5$ τ and $1 \times 10^6$ τ were used for the $N_{bb}$ = 80 and 240, respectively. The suitability of these simulation durations is supported by **Figure 3**(b) and (d), which illustrate the $\eta(t)$ profiles integrated from $G(t)$. For the case of $N_{bb}$ = 80, $\eta(t)$ plateaus by $2 \times 10^5$ τ, while for $N_{bb}$ = 240, convergence is achieved by around $6 \times 10^5$ τ. Case 12 (240p1φ10σ50s12) is an outlier, with $\eta(t)$ increasing over the full integration window. However, since Case 12 (240p1φ10σ50s12) already exhibits the highest zero-shear viscosity among the long-chain systems, extending the simulation time would further increase $\eta_0$ without meaningfully impacting the relative trends or statistical analyses. Therefore, we conservatively set the cutoff integration time for long polymers at $1 \times 10^6$ τ.

### 3.1.2 Zero-shear viscosity calculation via Einstein method

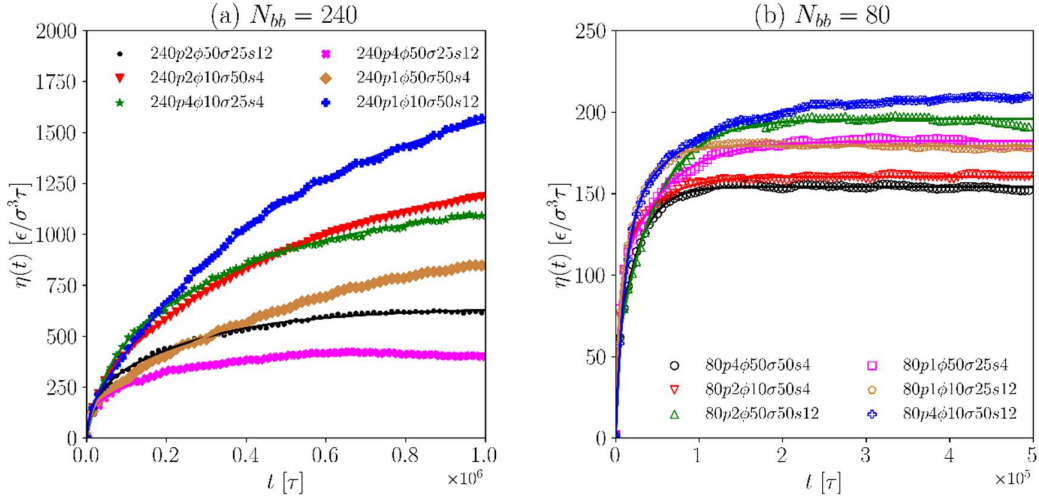

Figure 4: $\eta(t)$ calculations via the Einstein method in conjunction with the fitted double exponential function (solid lines). (a) $\eta(t)$ for long polymers ($N_{bb}$ = 240) based on the Einstein relation. (b) $\eta(t)$ for the short polymers ($N_{bb}$ = 80) based on the Einstein relation.

In the Einstein method (**Equation 4**), the integration $\frac{V}{2tk_\mathrm{B}T}\langle\left(\int_0^t \sigma_{ij}(t')dt'\right)^2\rangle$ for all short polymers climbs quickly to the plateau after $10^5$ $\tau$, as shown in **Figure 4(b).** In contrast, the integration of longer polymers rises more gradually, with a less distinct plateau emerging only after $8 \times 10^5$ $\tau$. To quantitatively describe the $\eta(t)$ profiles obtained from both the Green–Kubo and Einstein methods, we fitted the data using double exponential functions as described in **Equation 5**. The resulting fits are shown in **Figure 3** (Green–Kubo) and **Figure 4** (Einstein), respectively. We summarize the fitting zero-shear viscosity $\eta_0$ in **Figure 5**. The zero-shear viscosity obtained from the Green-Kubo method shows a strong agreement with the results from the Einstein method.

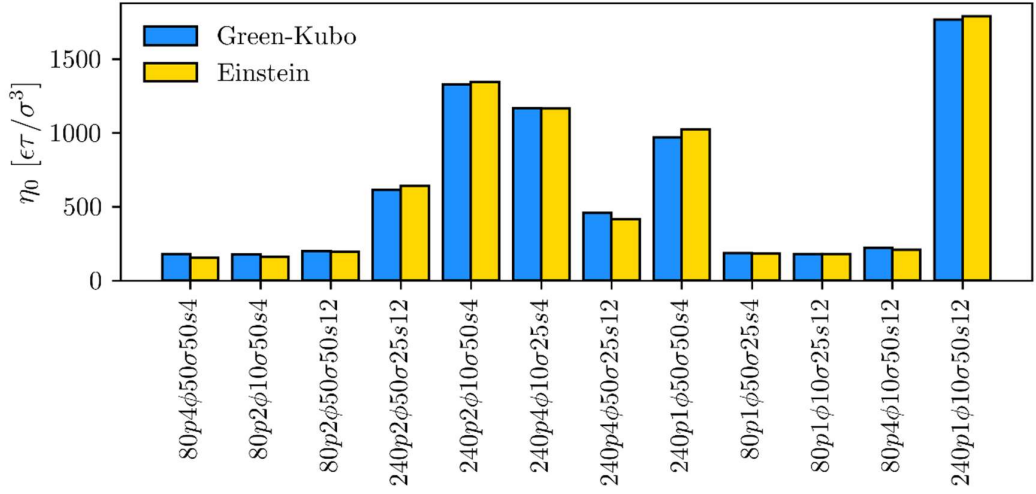

Figure 5: Comparison of $\eta_0$ from Green-Kubo (blue) and Einstein (gold) approaches.

*3.2 Rouse mode analysis of polymer relaxation*

To evaluate the influence of branching topology on backbone dynamics, we analyzed the normalized Rouse mode autocorrelation functions with the KWW stretched exponential model. The stretched exponential factor $\beta_p$ and effective relaxation time $\tau_p$ are exhibited in **Figure 6**.

In **Figure 6**(a), cases 5 (240p2φ10σ50s4) and 6 (240p4φ10σ25s4) exhibit minimal sidechain effects on backbone dynamics with the lowest composition factor ($\varphi = 0.1$) and shortest sidechains ($N_{sc} = 4$). Similar to linear polymers, their stretching exponential factor $\beta_p$ shows a minimum near $N/p = N_e$ and recovers gradually toward 0.8 as $N/p$ decreases gradually below $N_e$, indicating that interchain entanglement remains the dominant factor for relaxation. In contrast, case 4 (240p2φ50σ25s12), case 7 (240p4φ50σ25s12), and case 8 (240p1φ50σ50s4), all with longer grafted blocks, $\varphi = 0.5$, display markedly different $\beta_p$ behaviors. In cases 4 (240p2φ50σ25s12) and 7 (240p4φ50σ25s12), $\beta_p$ begins to deviate significantly from the short-grafted systems as $N/p$ falls below 30 - approximately the length of one grafted block. And this deviation continues until $N/p$ approaches the sidechain length $N_{sc}$. Case 8 (240p1φ50σ50s4) shows less deviation for $N/p > 60$ but exhibits the most pronounced suppression of $\beta_p$ at the smallest $N/p$, indicating enhanced local dynamic constraints by long sidechains.

**Figure 6**(c) and (d) present the effective relaxation time $\tau_p$ for the long polymers and the short polymers respectively. For $N/p > 10$, polymer relaxation is predominantly governed by the overall sidechain mass ($\varphi \times N_{sc} \times \sigma_{sc}$)- as the number and length of sidechains increase, the structural inertia rises, thereby slowing backbone relaxation. However, as $N/p$ approaches the sidechain length $N_{sc}$, the influence of the composition factor $\varphi$ diminishes.

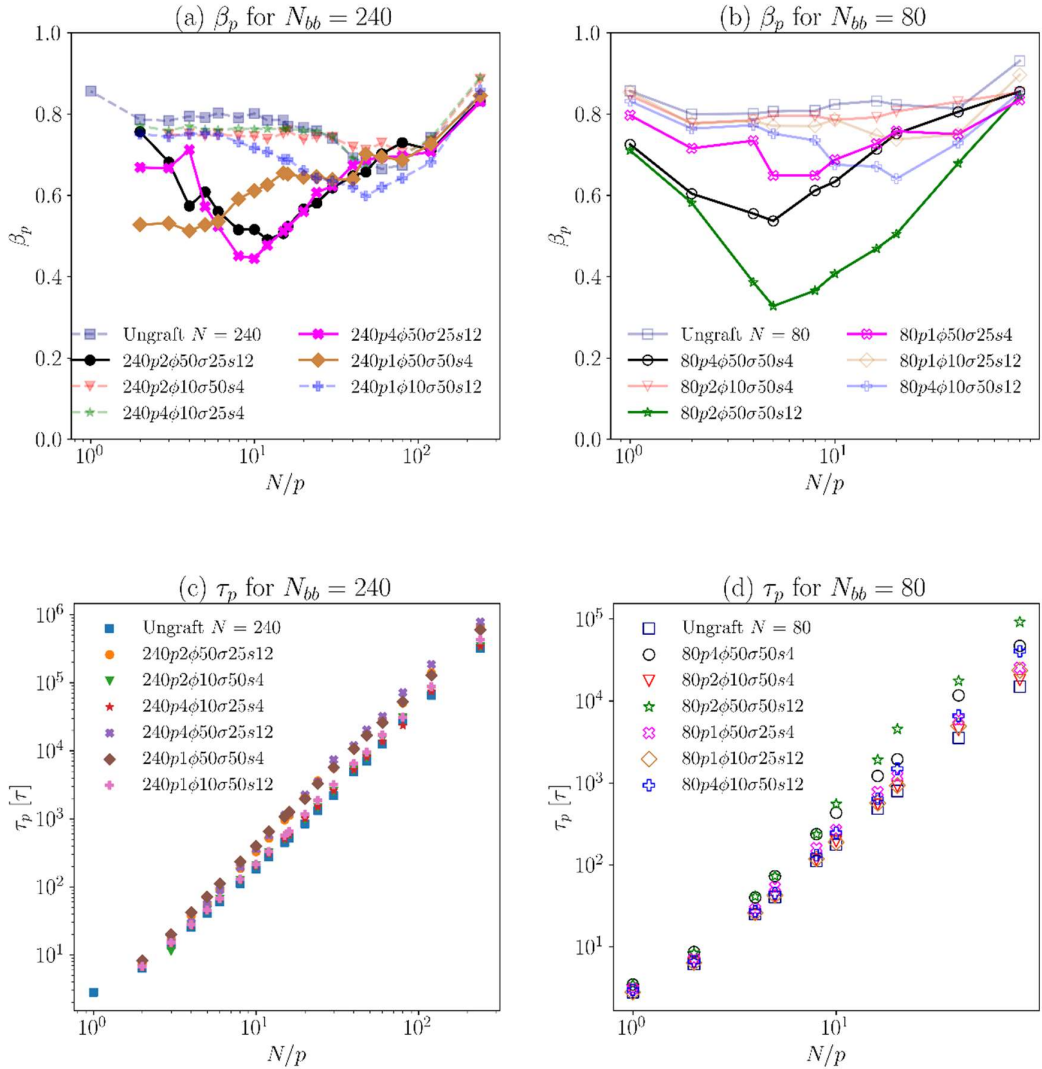

Figure 6. (a) Stretched exponential factor $\beta_p$ for $N_{bb} = 240$. (b) Stretched exponential factor $\beta_p$ for $N_{bb} = 80$. (c) Effective relaxation time $\tau_p$ for $N_{bb} = 240$. (d) Effective relaxation time $\tau_p$ for $N_{bb} = 80$. Please refer to Figure 1 regarding the topological details.

In the KWW model, the deviation of $\beta_p$ quantifies the deviation of the relaxation process from single-mode (purely exponential) behavior.[40] For example, an untangled polymer typically exhibits exponential relaxation ($\beta_p = 1$), whereas an entangled linear polymer displays deviations due to topological constraints, resulting in a minimum $\beta_p$ near $N/p = N_e$. In our simulated systems, we observe significant deviations from the behavior of ungrafted linear polymers in both short and long chains, indicating the presence of multimodal relaxation dynamics introduced by sidechains. To further investigate this, we analyzed the mean-squared displacement (MSD) using **Equation 12**:

$$\text{MSD}(t) = \frac{1}{N} \sum_{i=1}^{N} [\boldsymbol{r}_i(t) - \boldsymbol{r}_i(0)]^2 \tag{12}$$

where $N$ is the number of calculated CG beads. The results are displayed in **Figure 7**.

In both short and long polymers, the MSD of beads within the ungrafted blocks are quite similar, though slight deviations appear after $10^5$ τ for the long polymers and $10^4$ τ for the short polymers. This may suggest that, for weakly entangled polymers, grafted segments exert limited influence on the relaxation of ungrafted blocks. In contrast, the MSD($t$) within grafted segments behave remarkably differently. The differences are solely governed by grafting density or sidechain length. For instance, among the cases of long polymer, cases 8 (240p1φ50σ50s4) and case 12 (240p1φ10σ50s12) show the lowest MSD before 1,000τ. However, as time progresses, case 7 (240p4φ50σ25s12) exhibits a crossover behavior, eventually matching case 12 (240p1φ10σ50s12) in exhibiting the slowest dynamics.

The profound contrast between the grafted and ungrafted block dynamics explains significant deviation of $β_p$ near $N/p = N_{sc}$ observed in the polymer with composition factor $φ = 50\%$. Since the influence of sidechains is localized, they primarily affect backbone segments comparable in length to the sidechains themselves. As a result, near $N/p = N_{sc}$, KWW stretched exponential function struggles to accurately fit the relaxation behavior of polymers exhibiting a strong MSD disparity. This limitation arises because the KWW model assumes a homogeneous relaxation process and is inherently unable to reconcile two distinct, coexisting relaxation modes - one dominated by fast-moving ungrafted segments and one by slower, more constrained grafted regions.

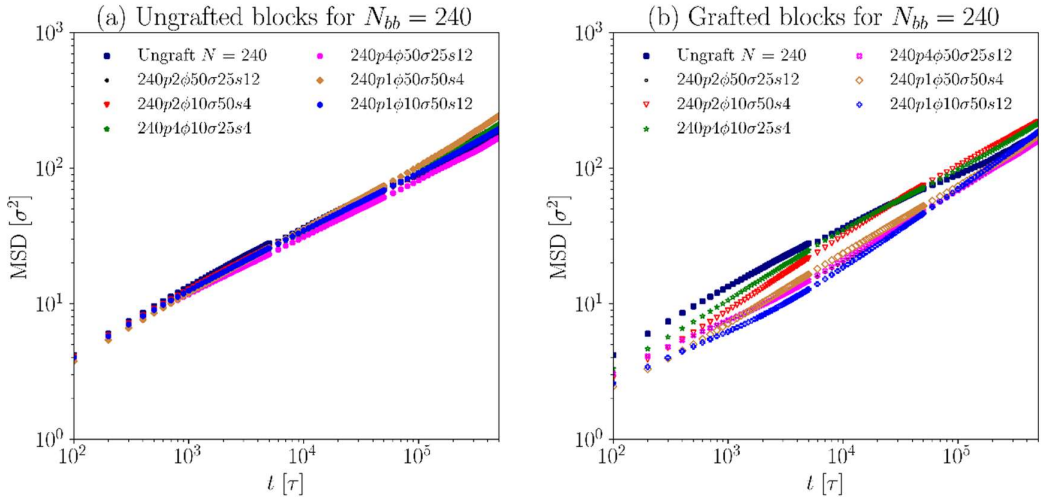

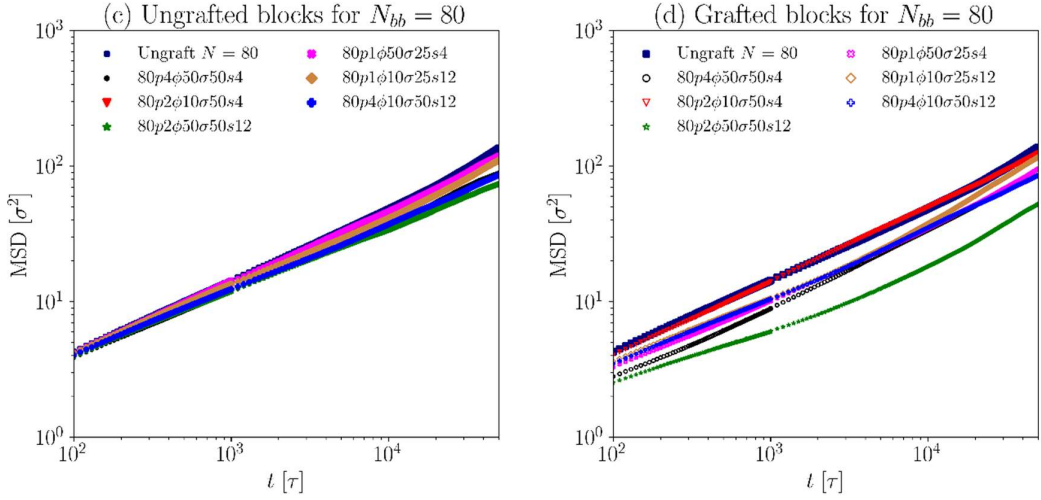

Figure 7: Monomeric MSD in the ungrafted and grafted blocks. The monomeric MSD of linear polymers with two ends are also included for comparison. For $N_{bb}$ = 240 in (a) and (b), we use MD trajectories spanning 0 to $10^4$ τ, $10^5$ τ, $10^6$ τ to compute the MSD over the ranges $0 \sim 5 \times 10^3$ τ, $5 \times 10^3$ τ $\sim 5 \times 10^4$ τ, and $5 \times 10^4$ τ $\sim 5 \times 10^5$ τ, respectively. For $N_{bb}$ = 80 in (c) and (d), we use MD trajectories from $0 \sim 2 \times 10^3$ τ, and $0 \sim 10^5$ τ to calculate the MSD over $0 \sim 10^3$ τ, and $0 \sim 5 \times 10^4$ τ, respectively

To more accurately capture the heterogeneous relaxation behaviors imposed by the grafted blocks, a refined KWW functional form is employed. This refined KWW introduces an additional fast relaxation mode, as detailed in **Equation 13**:[41-44]

$$\frac{\langle X_p(t)X_p(0)\rangle}{\langle X_p(0)X_p(0)\rangle} = [\alpha + (1-\alpha)e^{-(t/\tau_{f,p})}]e^{-(t/\tau_{s,p})^{\beta_p}}, \quad (13)$$

where $\tau_{f,p}$ and $\tau_{s,p}$ denote the fast and slow relaxation of mode $p$, respectively, and $\alpha$ controls the weighting between the two.

We compare the performance of original KWW (**Equation 7**) and the modified KWW model (**Equation 7**) for polymers with $\varphi = 0.5$ at $N/p = N_{sc}$ in **Figure 8**. The modified KWW functional form significantly improves the characterization of heterogeneous relaxation dynamics, especially during the early stages. This improvement suggests that the modified KWW can capture the slow relaxation mode overlooked by the original KWW, which assumes homogeneous relaxation. The fitted value of $\alpha$ (0.4 ~ 0.6) centers around 0.5, indicating that both fast and slow modes contribute comparably to relaxation.

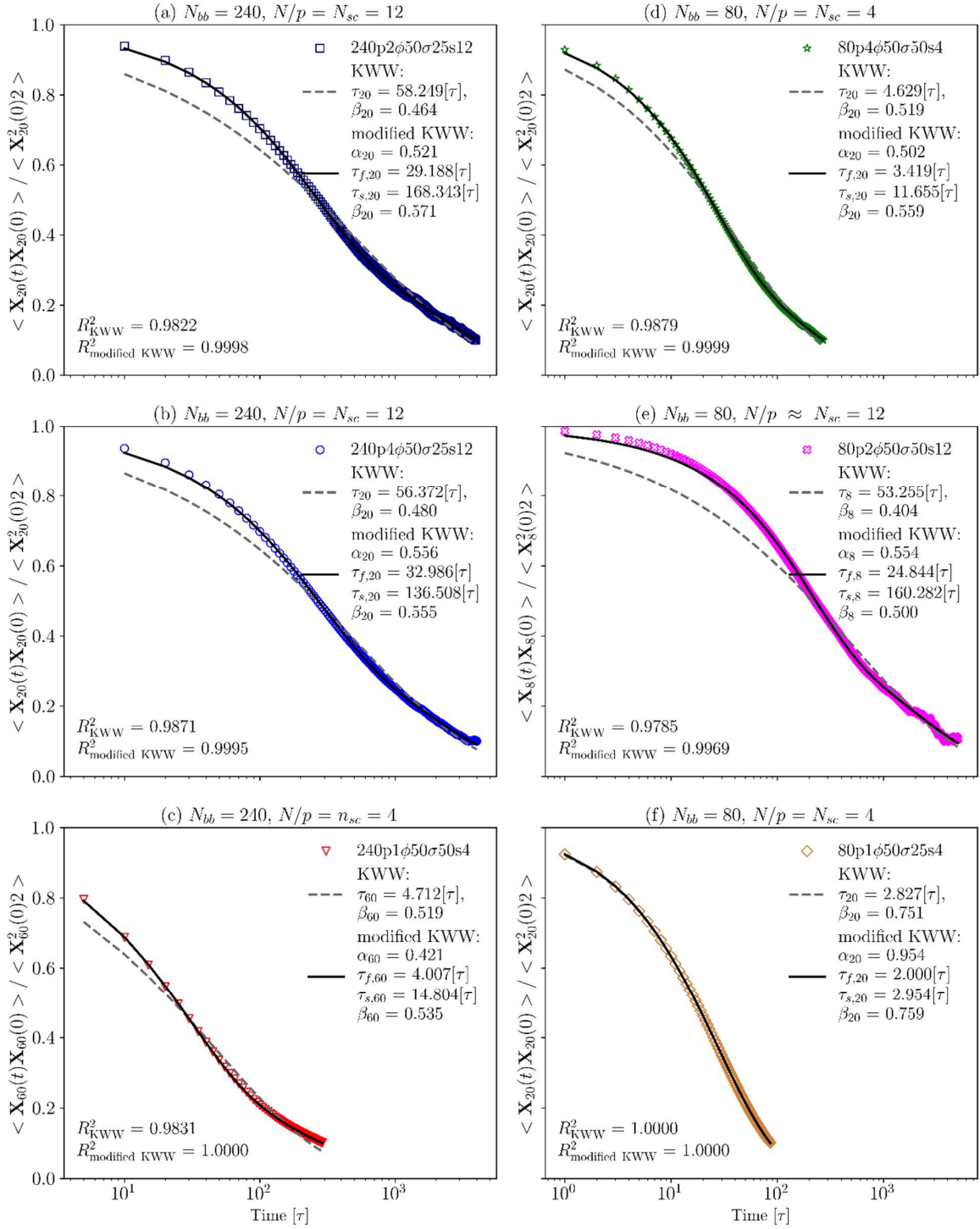

Figure 8: Comparison of the modified KWW and original KWW functionals on fitting the normalized Rouse mode autocorrelation. The coefficient of determination $R^2$ for the KWW and modified KWW fitting are compared in the lower left corner.

*3.3 CFD analysis of polymer printability.*

Since all properties obtained in MD simulations are expressed in dimensionless LJ units, we convert the zero-shear viscosity and relaxation time to physical units with scaling factors provided in **Table 3**.[45] The corresponding converted zero-shear viscosity and relaxation time are presented in **Figure 9**.

Table 2: Conversion of LJ units to physical units. The estimated Reynolds number Re = $\rho Ul/\eta$, where $\rho$, $U$, $l$, and $\eta$ refer to the density, velocity, nozzle diameter and apparent viscosity, respectively.

| Time | Viscosity | $T_{ref}$ | est. Re |
|---|---|---|---|
| 1 $\tau$ = 0.01 ns | 1 $\varepsilon/\tau\sigma^3$ = 0.176 mPa·s | 298 K | 31 ~ 332 |

For each polymer configuration, we perform CFD simulations for 3 seconds prior to data collection to ensure the system reaches steady-state flow, corresponding to approximately two full flow-through cycles. **Figure 9** presents a comparison of the pressure profiles at the nozzle inlet.

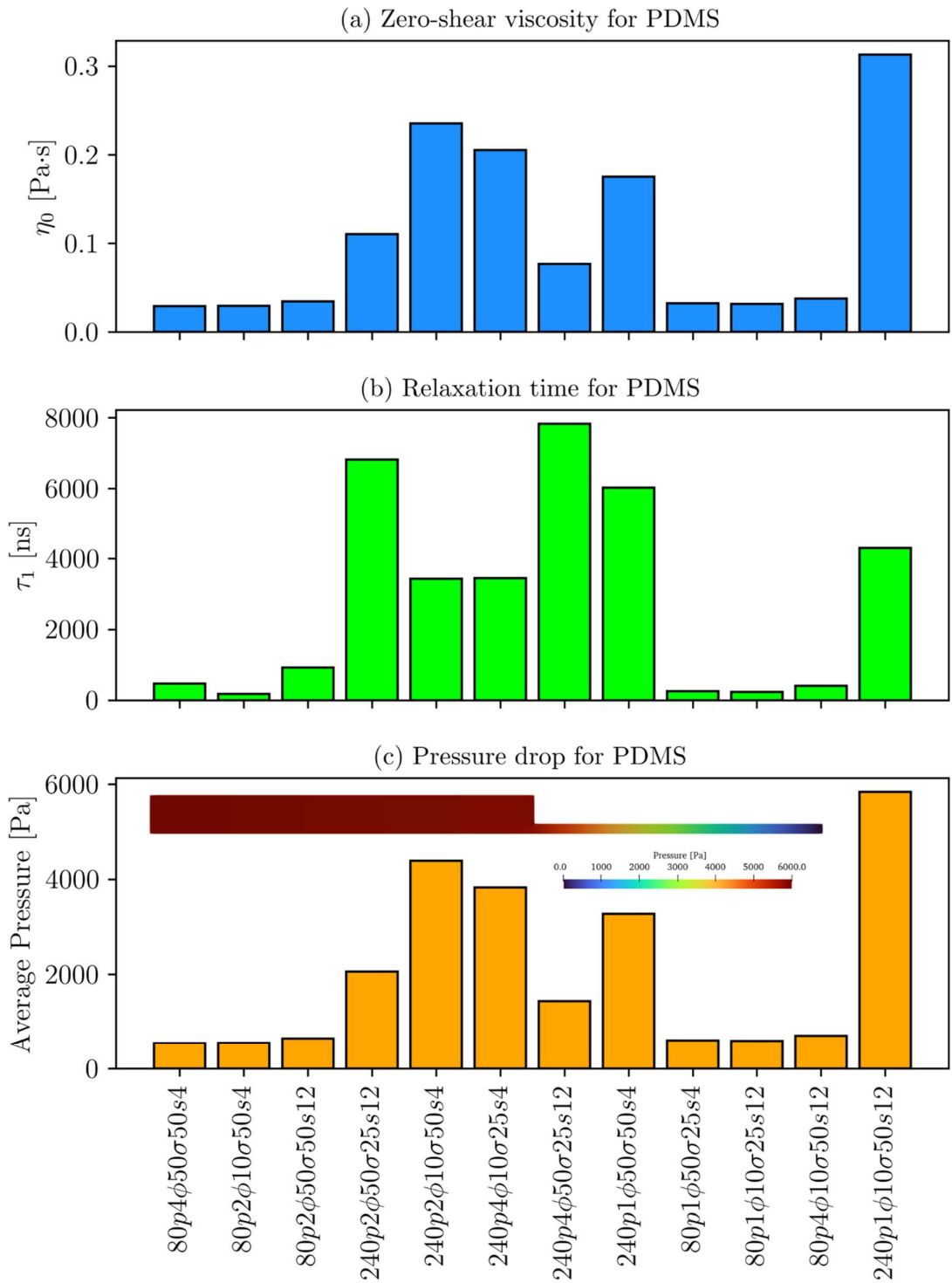

Figure 9: Comparison of (a) zero-shear viscosity for PDMS, (b) relaxation time for PDMS, and (c) pressure drop through the nozzle.

*3.4 statistical analysis of zero-shear viscosity, relaxation, and extrudability.*

In this section, we performed linear regression analysis to evaluate the relationships between simulated zero-shear viscosity, relaxation time, pressure difference and the topological parameters. The cross-validation for linear regression is shown in **Figure 10**. A correlation is considered positive when the regression coefficient is greater than zero, and negative when it is less than zero. Correlations with *p*-values greater than 0.05 (commonly considered statistically insignificant) are considered to not be distinguishable from zero. The results are summarized in **Table 3**. Within the scope of our designed polymers (untangled or weakly entangled polymers), both zero-shear viscosity and pressure drop show a positive correlation with backbone length and a negative correlation with the length of the grafted block. This trend suggests that polymers with concentrated grafted blocks, such as pom-pom tree polymers, exhibit higher zero-shear viscosity compared to stochastically branched polymers, which hinders the extrudability. In addition to backbone length, both the sidechain length and a higher grafting ratio contribute to increasing relaxation time. This observation is supported by the monomeric (MSD) shown in **Figure 7**. Since the polymers in this study are not highly entangled, the effect of entanglement dilution from the sidechains is minimal. Instead, the inertia of the grafted sidechains hinders local backbone relaxation.

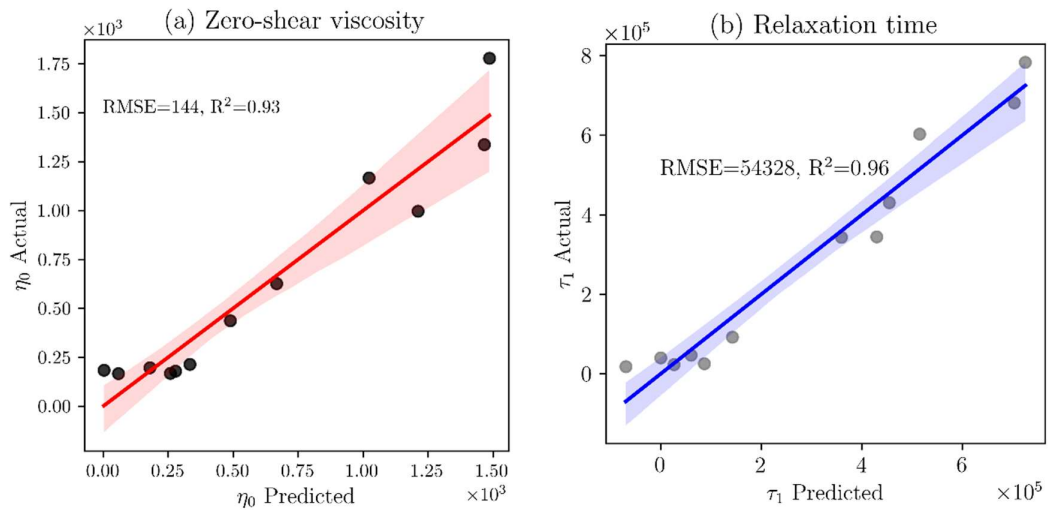

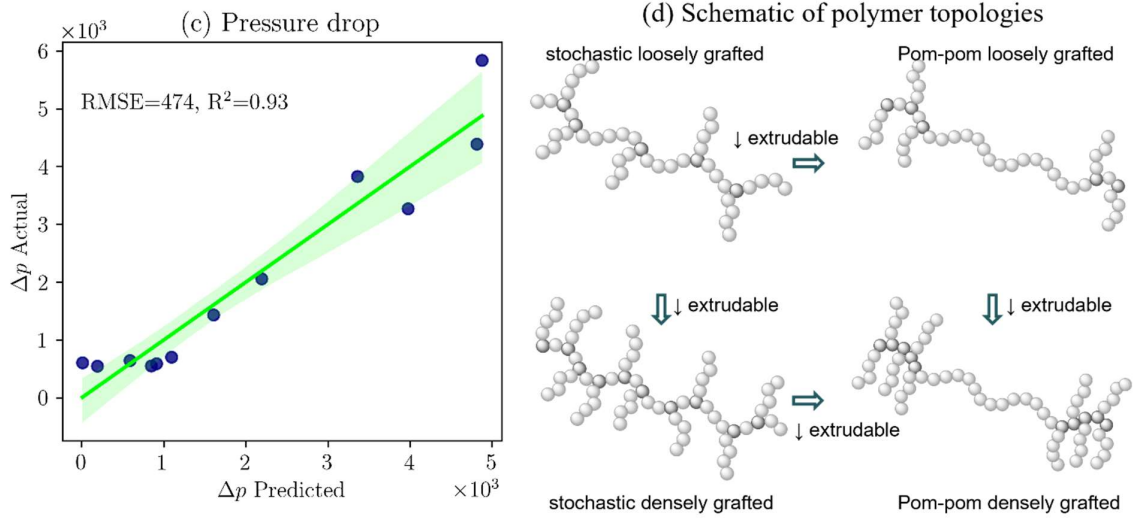

Figure 10: Statistical analysis of the effect of molecular topology on (a) zero-shear viscosity, (b) relaxation time, and (c) pressure drop along the nozzle. Detailed results are provided in Table (d). The shaded region represents the 95% confidence interval.

Table 3: Regressed effects on zero-shear viscosity, relaxation dynamics, and pressure drop. Effects with $p$-values greater than 0.05 are considered statistically insignificant and are marked as N.A. (shaded in dark).

|  | Zero-shear viscosity | Relaxation time | Pressure drops |
|---|---|---|---|
| Backbone length $N_{bb}$ | Positive | Positive | Positive |
| Periods $n_p$ |  |  |  |
| Composition factor $\varphi$ | Negative | Positive | Negative |
| Graft density $\sigma_{sc}$ | Positive |  | Positive |
| Sidechain length $N_{sc}$ |  | Positive |  |

## 4. Conclusions and outlooks

In this work, we developed a bottom-up, cross-scale strategy to quantify the influence of polymer topology on printability by integrating nanoscale CGMD simulations with continuum-scale CFD. The topological parameters examined include backbone length, sidechain length, grafting density, grafted block proportion, and the periodicity of grafted and ungrafted segments. For each polymer architecture, CGMD simulations were used to obtain zero-shear viscosity and
21

relaxation time, which served as input parameters for the PTT rheology model to simulate pressure drop during flow of PDMS melts through a nozzle.

We used linear regression to analyze the relationships between zero-shear viscosity, Rouse time, pressure drop, and the topological parameters listed above. The statistical analysis indicates that concentrated graft blocks negatively impact zero-shear viscosity and pressure drop, implying that polymers with concentrated graft blocks (e.g., pom-pom polymers) require higher extrusion pressures than stochastically branched polymers. In untangled or weakly entangled melts, Rouse time increases with both a higher grafted block ratio and longer sidechains, attributed to sidechain inertia.

## 5. Acknowledgement

This research was sponsored by the U.S. Department of Energy, Office of Science, Basic Energy Sciences, Materials Sciences and Engineering Division (FWP# ERKCK60), under Contract DEAC05-00OR22725 with UT-Battelle, LLC. Computational work was conducted as part of a user project at the Center for Nanophase Materials Sciences (CNMS), which is a US Department of Energy, Office of Science User Facility at Oak Ridge National Laboratory. This research used resources of the Oak Ridge Leadership Computing Facility at the Oak Ridge National Laboratory, which is supported by the Office of Science of the U.S. Department of Energy under Contract No. DE-AC05-00OR22725.

# Supplementary Information

# Cross-scale Modeling of Polymer Topology Impact on Extrudability through Molecular Dynamics and Computational Fluid Dynamics

*Yawei Gao[a], Jan Michael Carrillo[b], Logan T. Kearney[a], Polyxeni P. Angelopoulou[a], Nihal Kanbargi[a], Arit Das[a], Michael Toomey[a], Bobby Sumpter[b], Joshua T. Damron[a,\*], and Amit K Naskar[a,\*]*

[a]Chemical Sciences Division, Oak Ridge National Laboratory, 1 Bethel Valley Rd, Oak Ridge, TN 37830

[b]Center for Nanophase Materials Sciences, Oak Ridge National Laboratory, 1 Bethel Valley Rd, Oak Ridge, TN 37830

---

[*] Corresponding author.

*Email addresses*: damronjt@ornl.gov (Joshua T. Damron), naskarak@ornl.gov (Amit. K. Naskar)

*May 1, 2025*

(a) Case 1 (80p4φ50σ50s4)

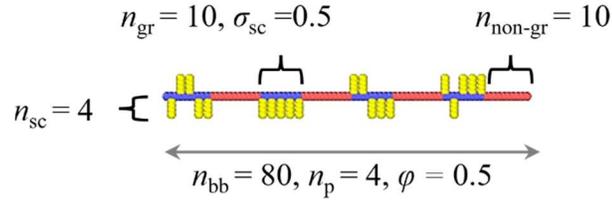

(b) Case 2 (80p2φ10σ50s4)

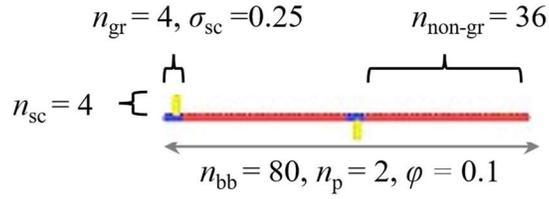

(c) Case 3 (80p2φ50σ50s12)

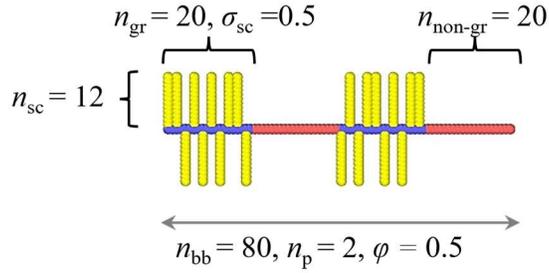

(d) Case 4 (240p2φ50σ25s12)

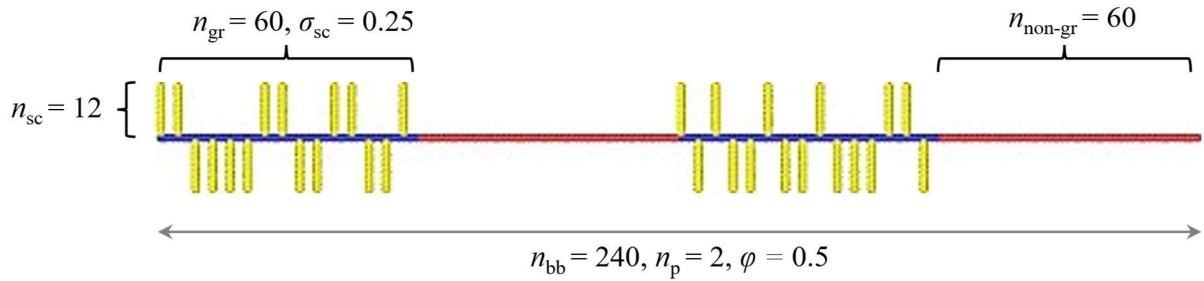

(e) Case 5 (240p2φ10σ50s4)

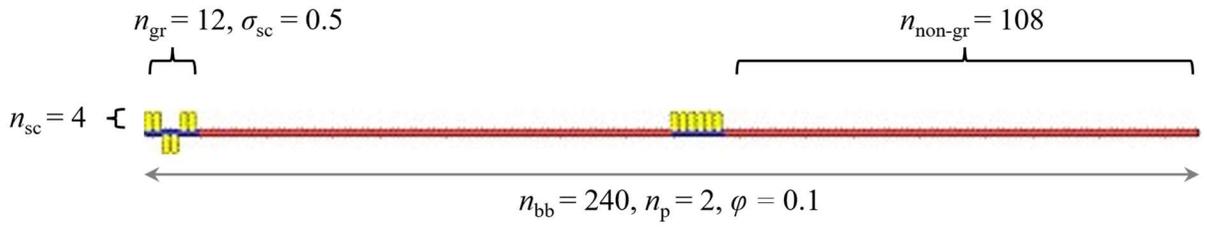

(f) Case 6 (240p4φ10σ25s4)

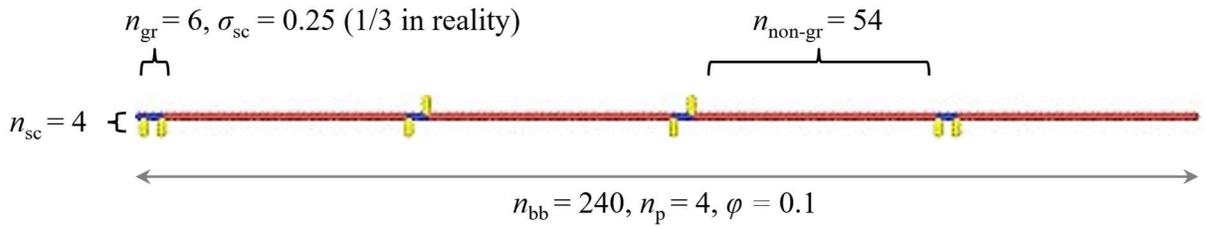

(g) Case 7 (240p4φ50σ25s12)

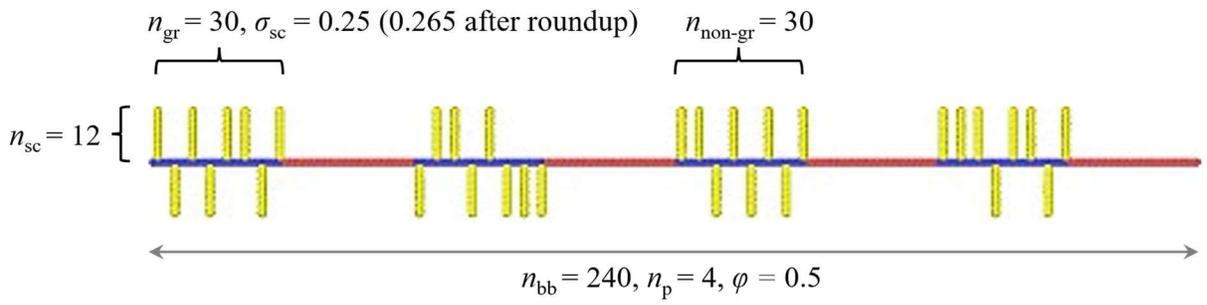

(h) Case 8 (240p1φ50σ50s4)

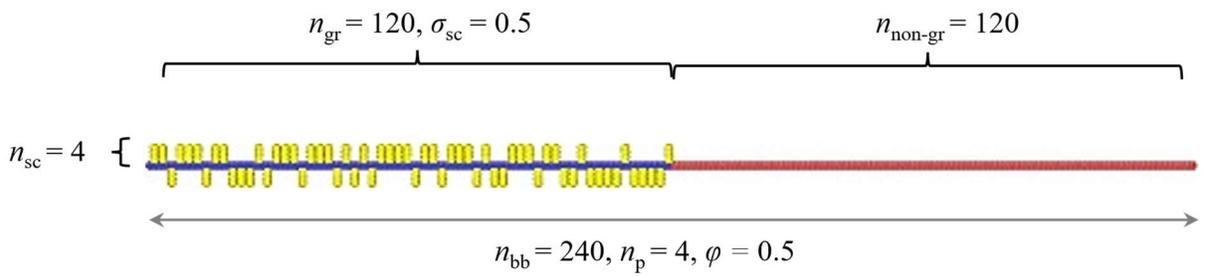

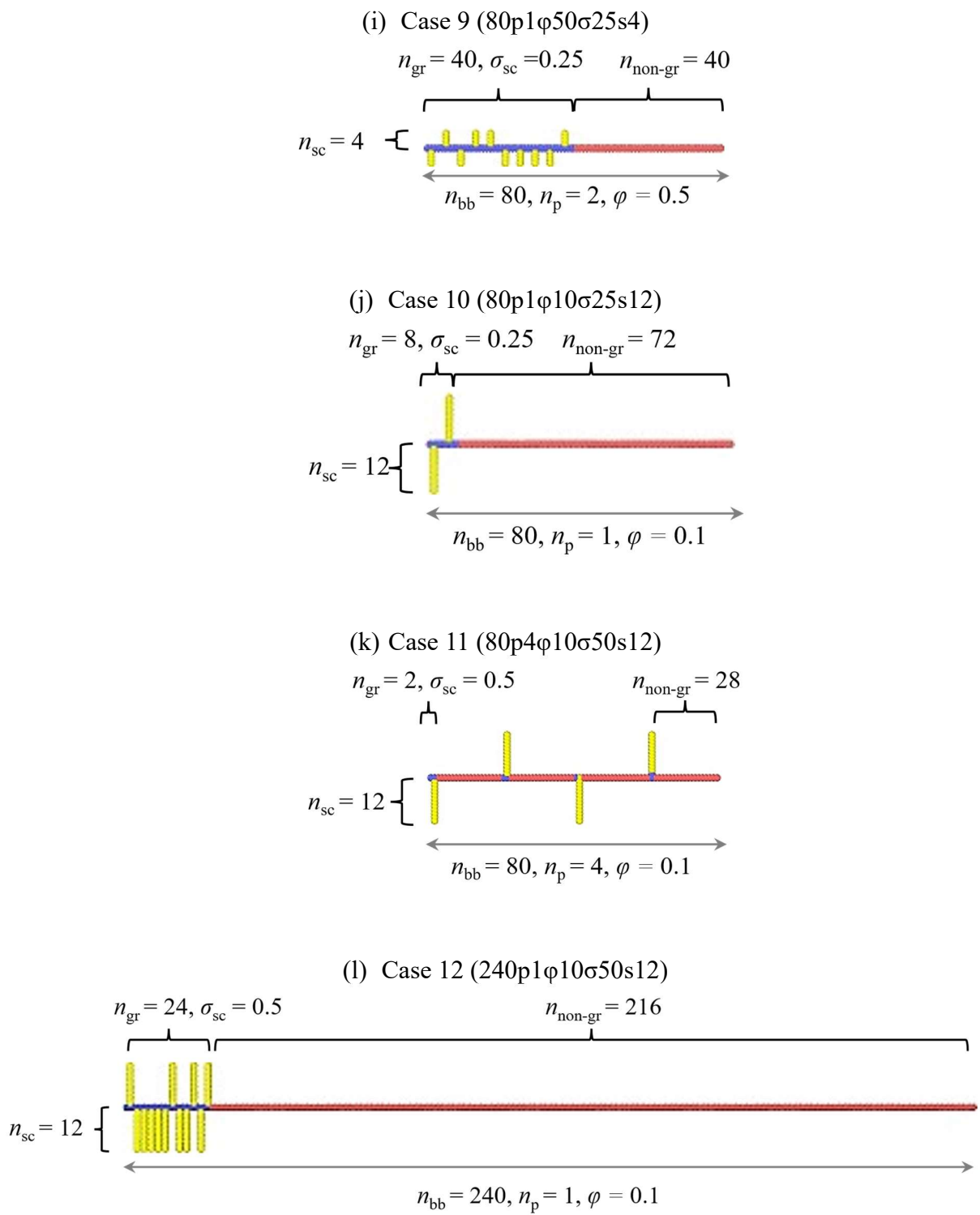

**Fig. S1| Polymer architecture.** Color scheme: grafted blocks (blue), sidechains (yellow), and nongrafted blocks (red)

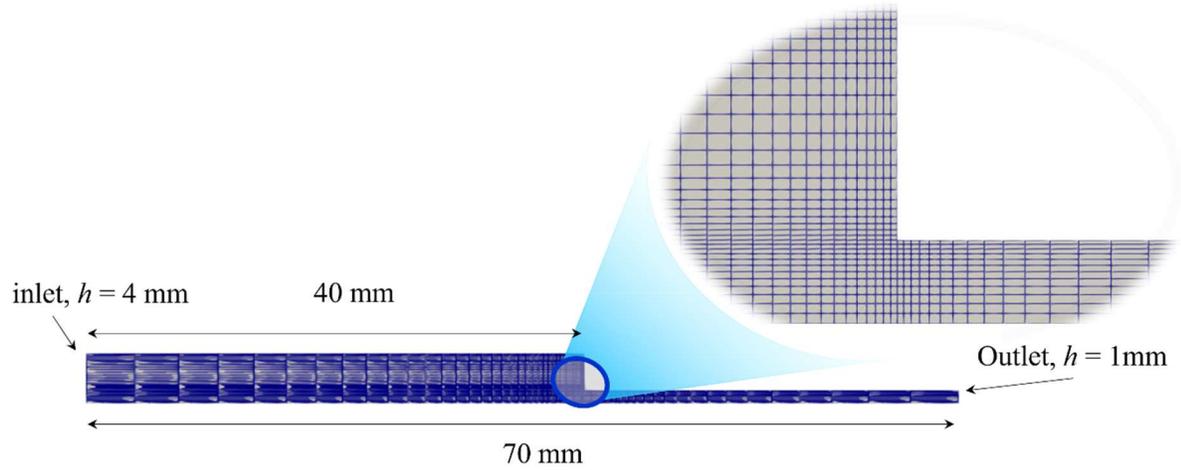

**Fig. S2| Flow domain model.** The simulation domain consists of 1,800 cells. A finer mesh with 14,400 cells was also tested, and the results remained consistent, confirming mesh independence.

**Table 1: Boundary conditions for CFD simulation.**

|  | Inlet | Outlet | Upper/lower walls | Front/Back boundaries |
|---|---|---|---|---|
| Velocity | 0.05 m/s | zeroGradient | noSlip | empty |
| Pressure | zeroGradient | 0 | zeroGradient | empty |
| Stress | zeroGradient | zeroGradient | zeroGradient | empty |

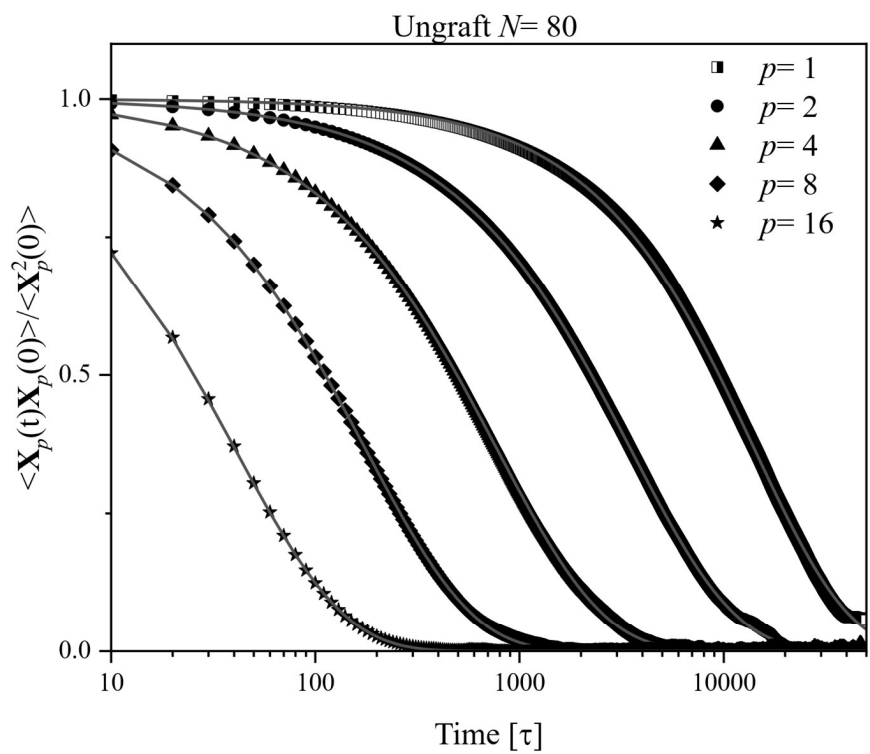

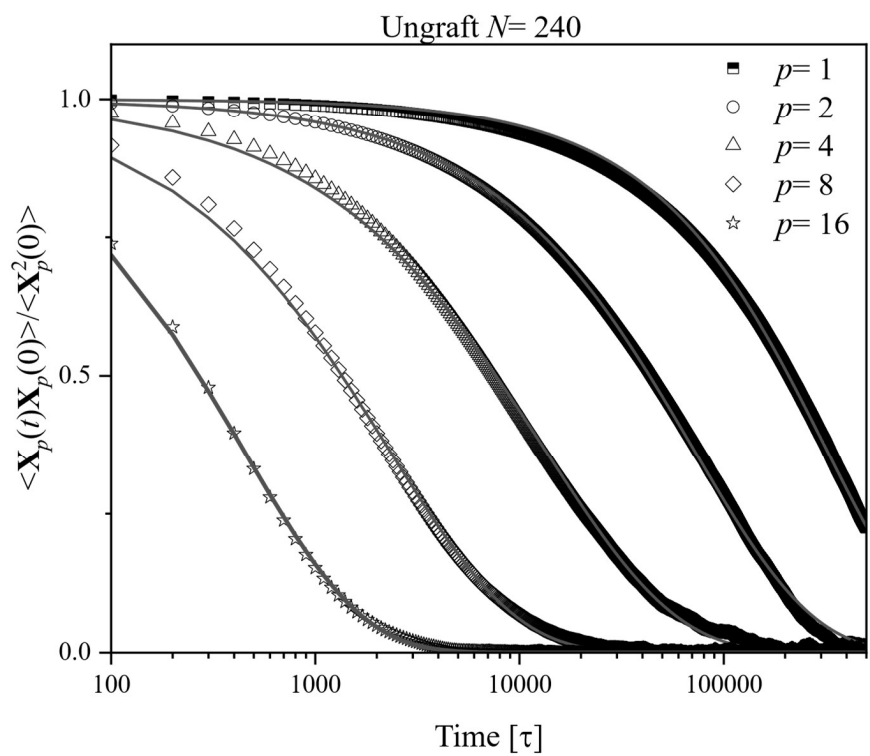

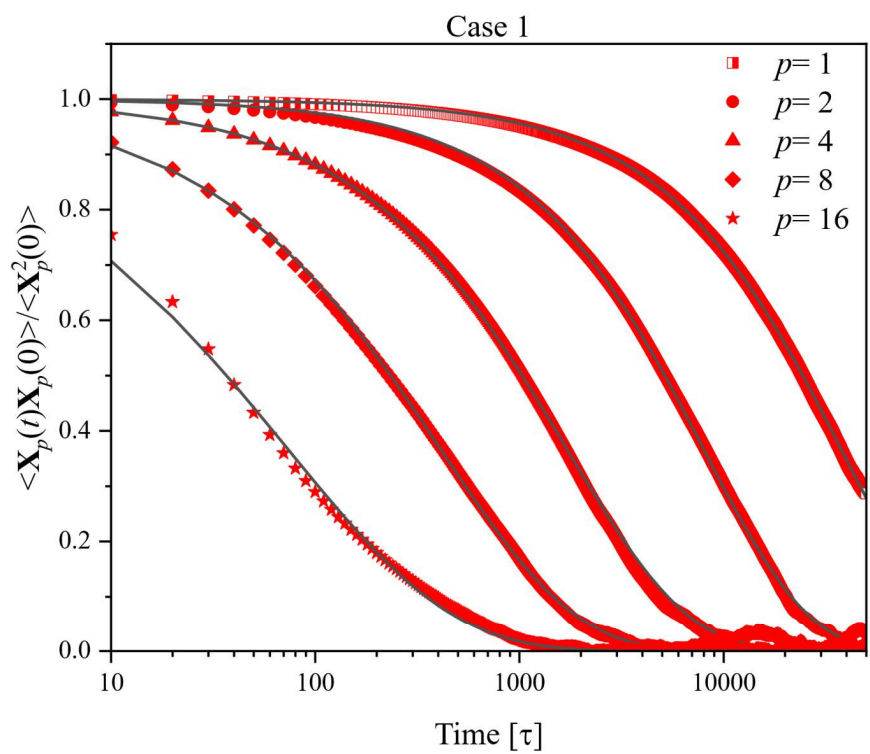
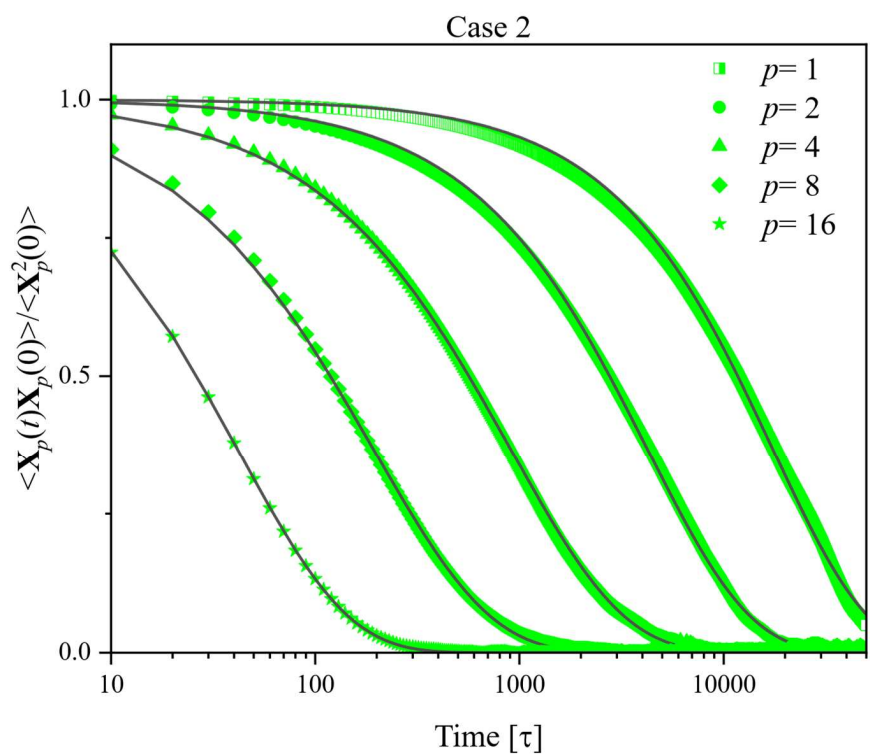

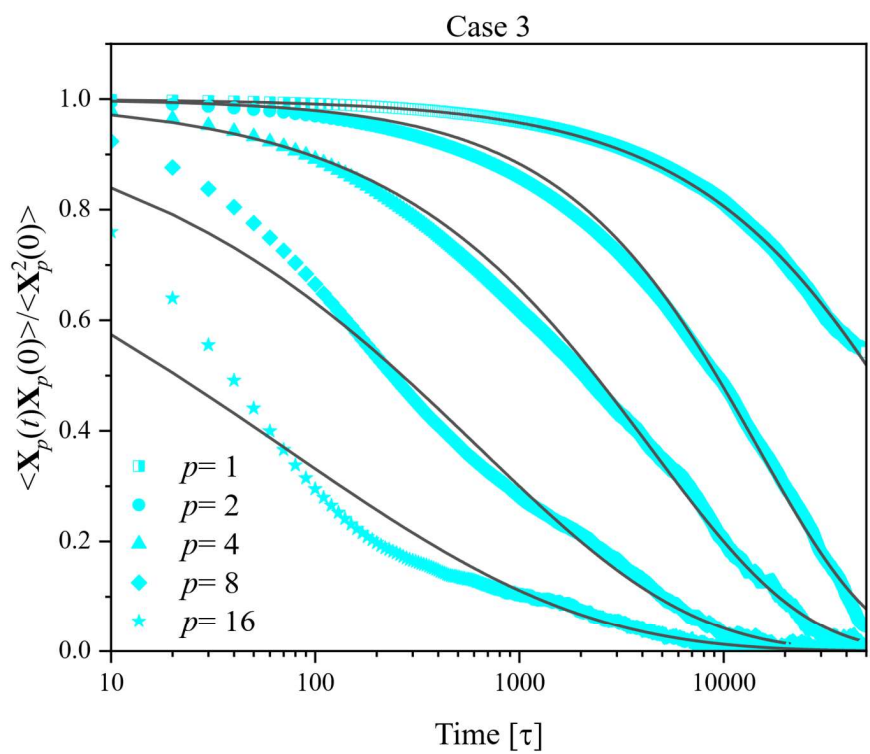
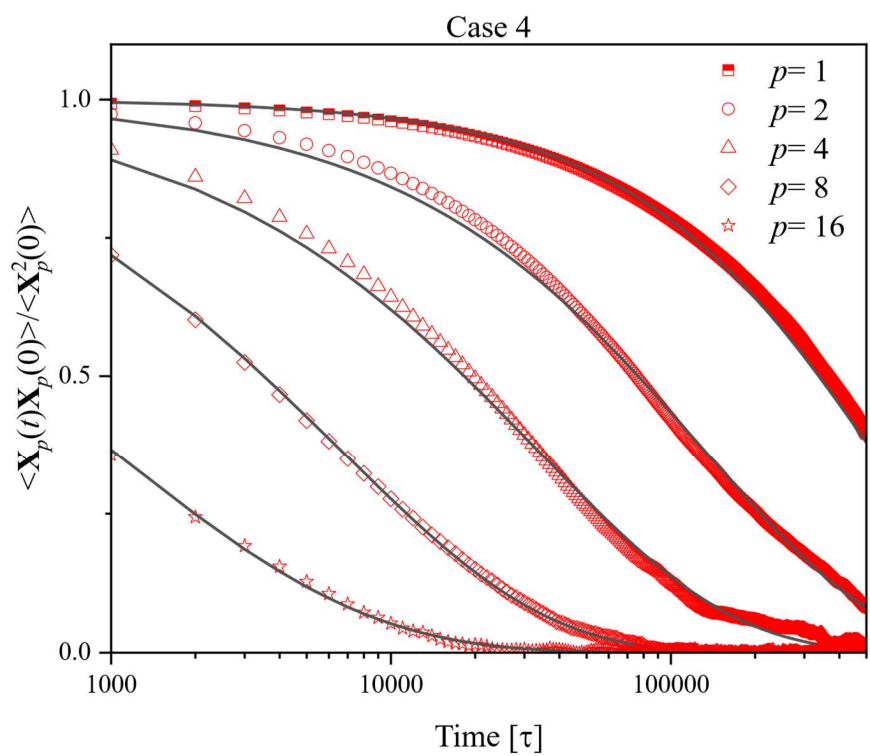

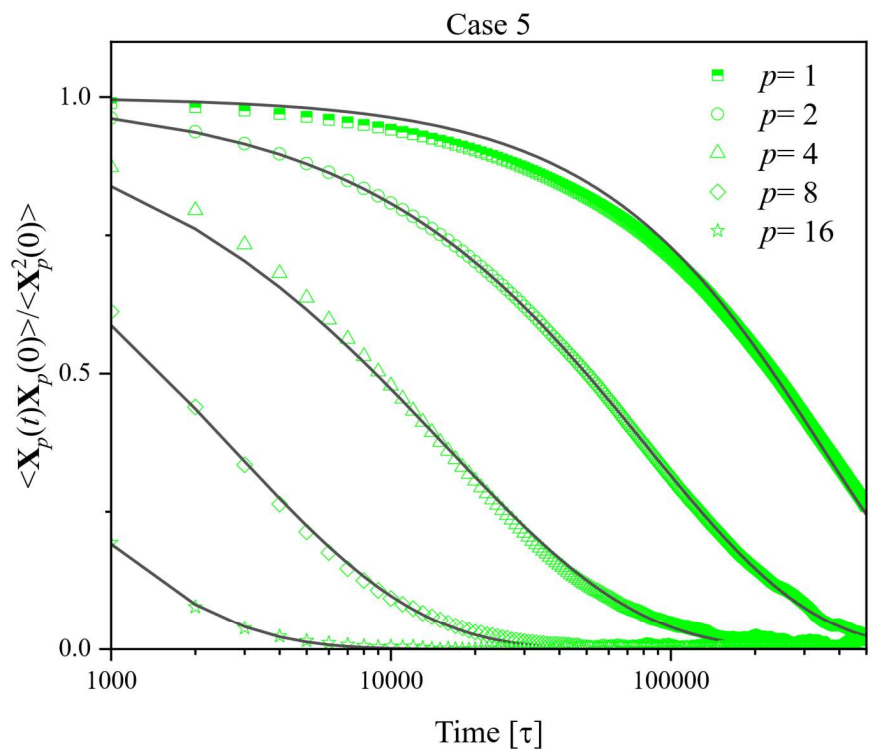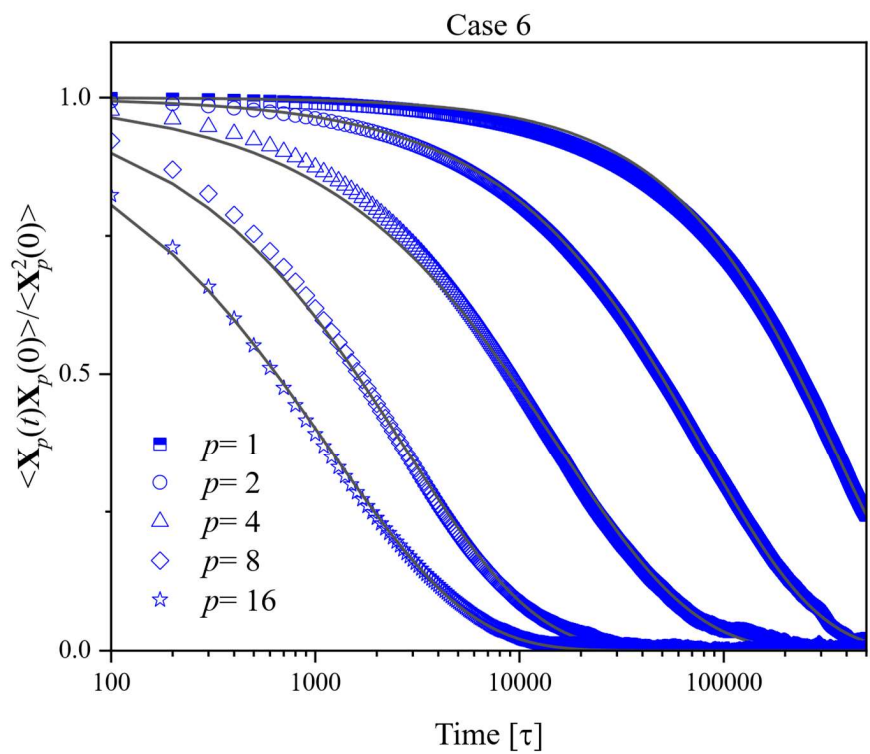

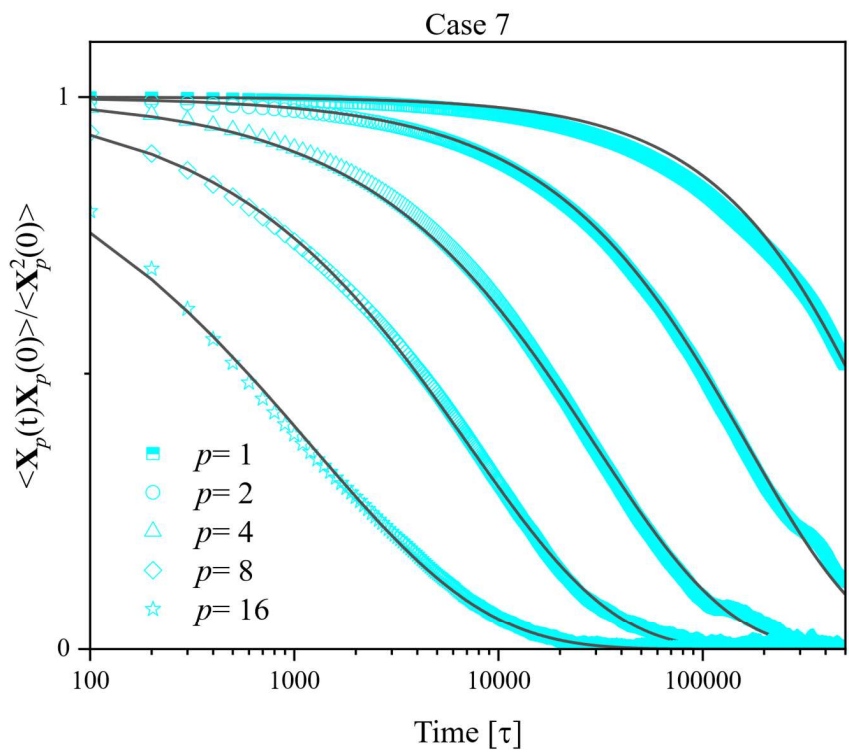

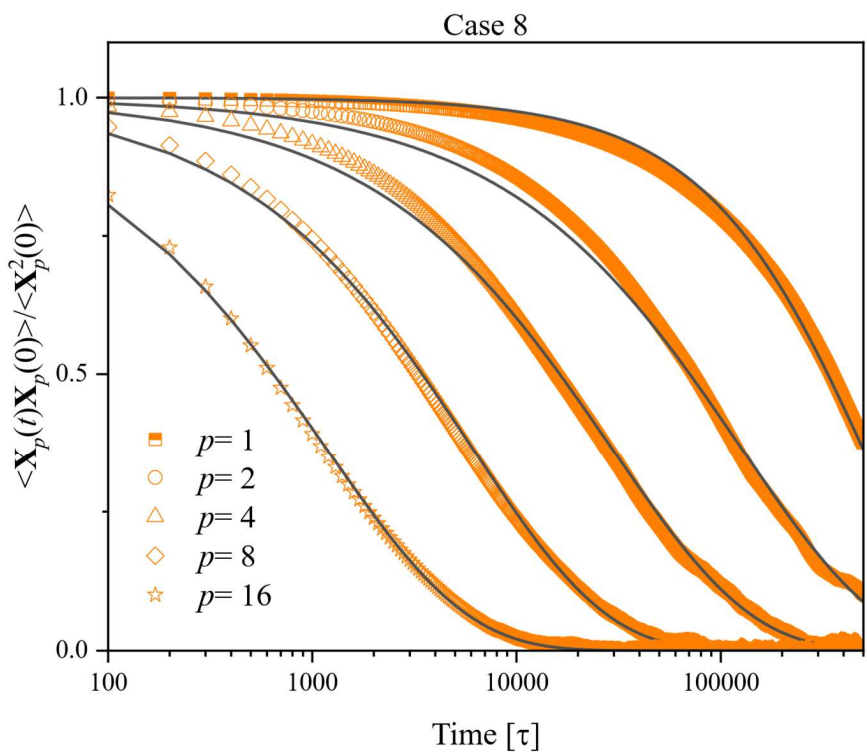

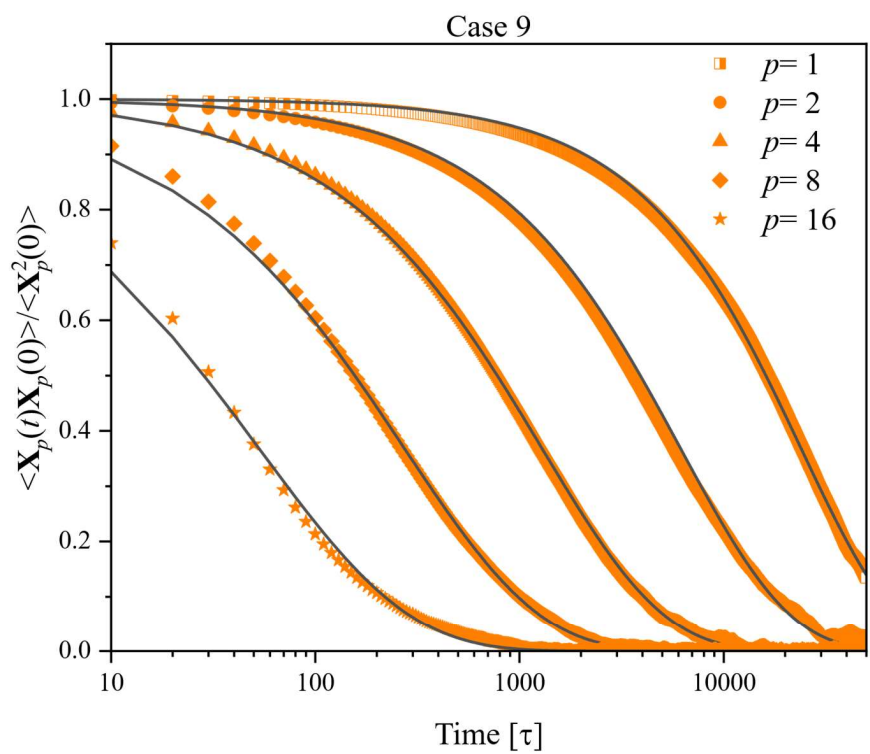

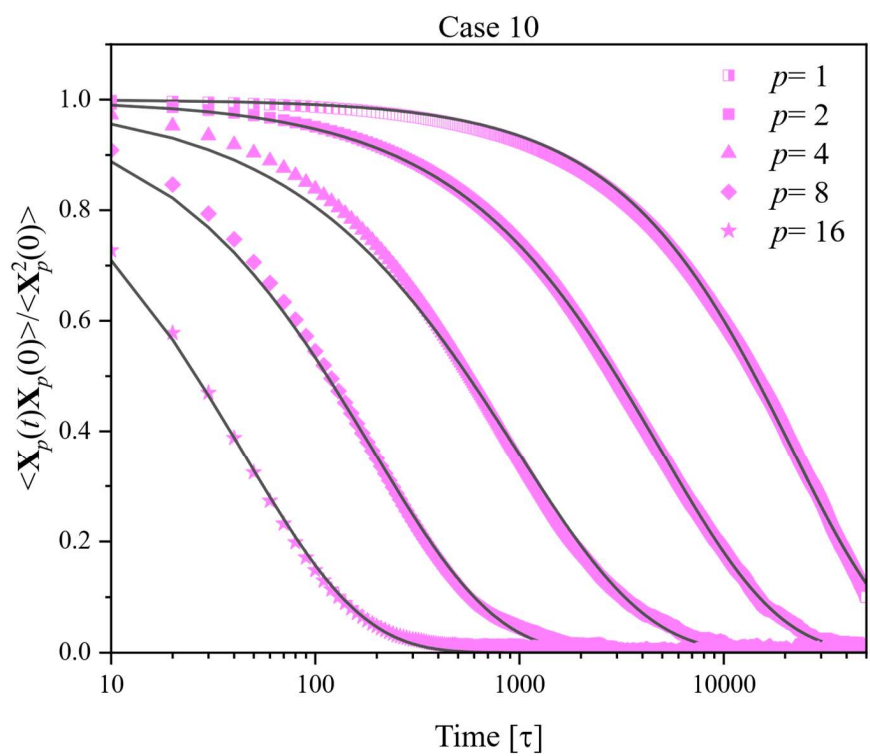

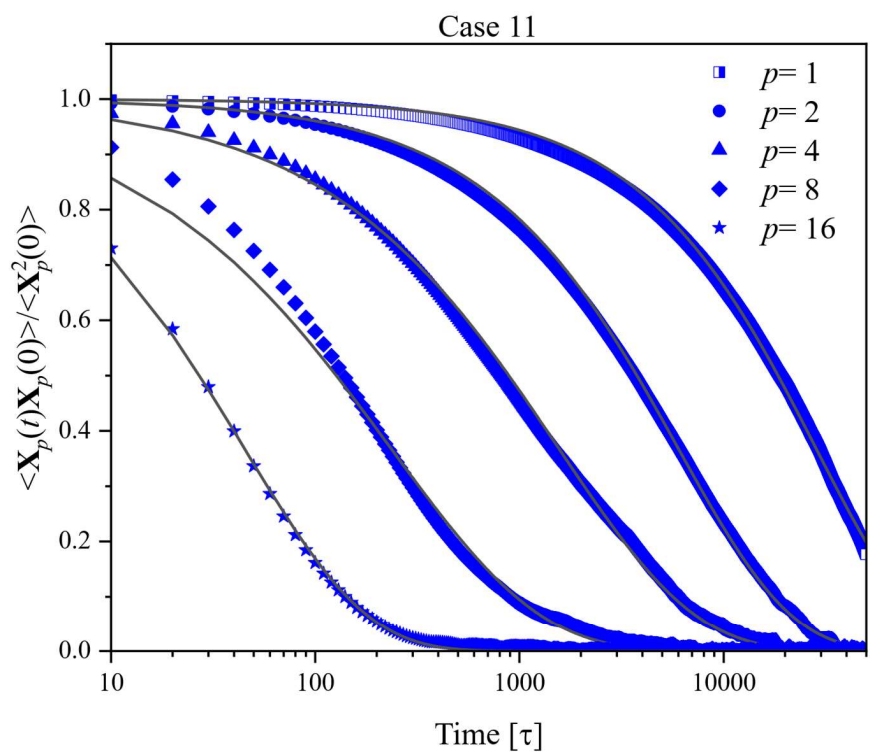
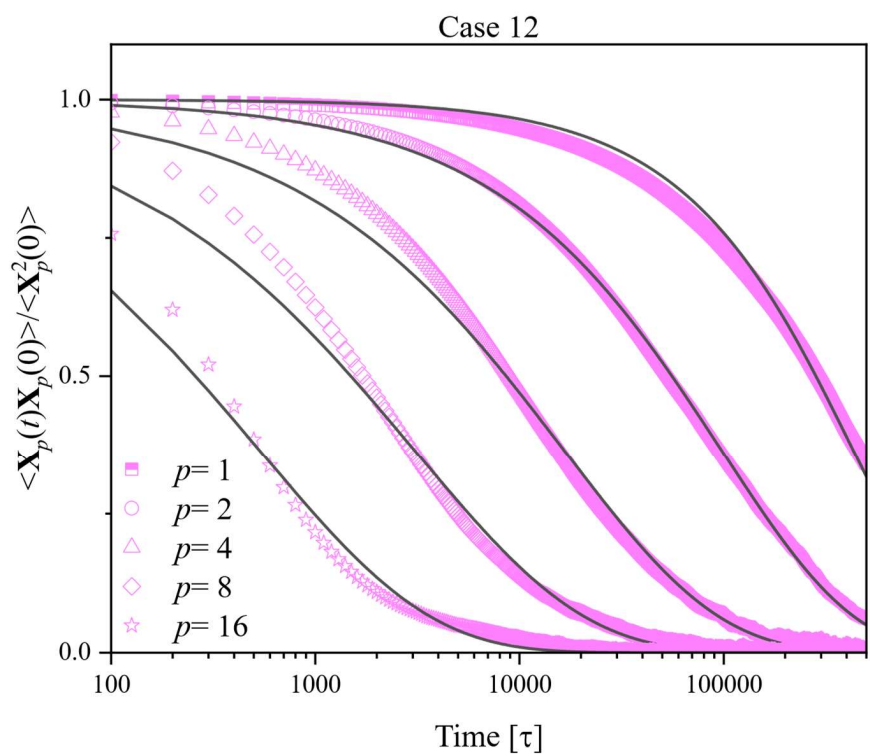

**Fig. S3| Normalized autocorrelation function.** We present the normalized autocorrelation function of Rouse modes $p$ = 1, 2, 4, 8, and 16 for all branching architectures in conjunction with their ungrafted linear counterparts. Besides, the normalized autocorrelation function is fitted with the KWW stretched exponential functions.